\documentclass[aps,prfluids,reprint,unsortedaddress,onecolumn,longbibliography,10.5pt]{revtex4-2}
\usepackage{hyperref}
\usepackage{lipsum}
\usepackage[caption=false]{subfig}
\usepackage{graphicx}
\usepackage[utf8]{inputenc}
\usepackage{amsmath}
\usepackage{amssymb}
\usepackage{graphicx}
\usepackage{siunitx}
\usepackage{mathtools}
\usepackage{booktabs}
\usepackage{placeins}

\newcommand{\ret}{\text{Re}_{\tau}}

\hypersetup{%
    pdfpagemode={UseOutlines},
    bookmarksopen,
    pdfstartview={FitH},
    colorlinks,
    linkcolor={black},
    citecolor={black},
    urlcolor={blue}
  }

\begin{document}


\title{Network analysis of Reynolds number scaling in wall-bounded Lagrangian mixing}
\author{Davide Perrone}
\email[]{davide.perrone@polito.it}
\affiliation{Department of Mechanical and Aerospace Engineering, Politecnico di Torino, 10129 Turin, Italy}
%
\author{J.G.M. Kuerten}
\affiliation{Department of Mechanical Engineering, Eindhoven University of Technology,
P.O. Box 513, 5600 MB Eindhoven, The Netherlands}

\author{Luca Ridolfi}
\affiliation{Department of Environmental, Land and Infrastructure Engineering, Politecnico di Torino, 10129 Turin, Italy}

\author{Stefania Scarsoglio}
\affiliation{Department of Mechanical and Aerospace Engineering, Politecnico di Torino, 10129 Turin, Italy}

\date{\today}

\begin{abstract}
The dispersion and mixing of passive particles in a turbulent channel flow is investigated by means of a network-based representation of their motion. We employ direct numerical simulations at five different Reynolds numbers, from $\ret = 180$ up to $\ret = 950$, and obtain sets of particle trajectories via numerical integration.
By dividing the channel domain into wall-normal levels, the motion of particles across these levels is used to build a time-varying complex network, which is able to capture the transient phase of the wall-normal mixing process and its dependence on the Reynolds number, $\ret$. 
Using network metrics, we observe that the dispersion of clouds of tracers depends highly on both their wall-normal starting position and the time elapsed from their release. We identify two main mechanisms that contribute to the long lasting inhibition of the dispersion of particles released near the walls. We also show how the relative importance of these mechanisms varies with the Reynolds number. In particular, at low $\ret$ the weaker velocity fluctuations appear dominant in inhibiting dispersion, while at higher Reynolds numbers a larger role is played by cyclic patterns of motion.
At the higher Reynolds numbers employed in this work, we find that most network properties are Reynolds-independent when scaled with outer flow variables. Instead, at lower $\ret$, the aforementioned scaling is not observed. We explore the meaning of the emergence of this scaling in relation to the features of dispersion and to the network definition.
\end{abstract}

\maketitle

\section{Introduction}

Dispersion and mixing of particles advected by turbulent flows occur in several fluid processes, including geophysical and environmental flows and many industrial applications \cite{dimotakis2005turbulent}. Turbulence has the ability to greatly enhance the dispersion of advected scalars; on the other hand, the complexity introduced by the multitude of scales present in high Reynolds number flows makes such problems difficult to analyze and describe. 
A straightforward, albeit simplified, representation for a passive species dispersed in a turbulent flow is the one provided by massless tracers which are simply advected by the underlying velocity field; in this context, the Lagrangian description of turbulent transport arises naturally \cite{toschi2009}. 
Past contributions have explored the statistical properties of advected tracers \cite{yeung89,wang,toschi05,xu06,stelz17}, particle pairs \cite{salazar2009two,buaria15} and larger structures \cite{biferale2005multiparticle,polanco18}, providing a characterization of the ensemble behavior of sets of tracers. Understanding these properties contributes to improve stochastic models for dispersion, that play a key role in practical applications \cite{sawford2001stocmodels,sawford1991pofafd}. 

In homogeneous isotropic turbulence, the statistical properties of tracers do not depend on their location; also, the properties of particle pairs and of higher order structures depend only on their mutual distance and not on their orientation. This is not true for inhomogeneous and anisotropic flows, where both the physical location of particles and the orientation of groups of them matter. In wall bounded flows, dispersion properties depend on the wall-normal $y$ coordinate. Wall-bounded flows are also anisotropic, so that wall shear enhances the dispersion of pairs whose separation lies in the wall normal direction; additionally, coherent structures of various sizes play a role in diversifying the mixing properties inside a wall-bounded flow \cite{smits, marusic_wb}. 
The importance of wall-bounded flows in nature and in industrial applications makes their mixing processes compelling, although one has to deal with the added complexity due to inhomogeneity and anisotropy. 

In recent years new methods based on networks, which are topological structures comprising a set of interacting objects (nodes) and a set of their mutual interactions (links), have proven useful in the analysis of fluid dynamics data \cite{iacobelloreview,donnerreview,sujith,yeh2021}. Classically, networks have been used to describe the interactions between, as an example, large groups of people, biological systems, electrical grids or transport infrastrucures, allowing to uncover the underlying collective behavior of the elementary units that constitute the network \cite{albert4947,newman2018networks,boccaletti}. 
Because of their ability to condense large systems in a simplified description, networks appear suitable to represent the complex dynamical interactions taking place in turbulent flows. The inherent complexity of fluids at high Reynolds numbers, along with the increase in quantity and resolution of fluid flow data (especially from high resolution simulations), are among the main reasons to find new processing and analysis methods. 

Several networks have been defined to extract specific information from flow fields; common applications include networks based on correlation of physical quantities (either fixed points in space \cite{scarsoglio2016complex} or trajectories \cite{schlueter2017coherent, vieweg}), the analysis of time-series \cite{charakopoulos,chara21}, the proximity of particle trajectories \cite{rypina2017npg,banisch} or vortical interactions \cite{taira2016jfm}. Another network-based method, \textit{i.e.} the \textit{transport network} formalism, has been proposed to describe the material connections between discrete regions of the fluid domain \cite{froyland2007prl,sergiac}. 
In particular, nodes are defined as the partitions of the domain and a link between them is established if the two regions have exchanged fluid in the time frame considered. This corresponds to a coarse-graining of the Lagrangian dynamics of the flow, which enables us to study the processes emerging from the collective behavior of fluid particles.
In previous works, our group has dealt with the properties of Lagrangian turbulence and scalar transport employing network-based methods. We demonstrated that the mapping of particle motion into geometrical objects is useful to unveil turbulent mixing features. \textcite{iacobello_plume} used visibility graph analysis to analyze the regularity and the occurrence of peaks in turbulent transport time-series in an experimental wall-bounded flow. \textcite{iacobello2019lagrangian}, instead, used a network accounting for the proximity of Lagrangian tracers to analyze the transition from an unmixed to a fully mixed state. Finally, \textcite{perrone20} used the same method employed in this work to provide results, at a fixed $\ret$, about the temporal evolution of particle dispersion and the existence of regions of the channel that have a reduced exchange of fluid between them. In this paper, we study a new aspect by focusing on the scaling properties of turbulent dispersion with respect to the Reynolds number $\ret$ in the range $\left[180,\,950\right]$; to achieve statistical significance of the Reynolds number comparison, we employ a much larger number of particle trajectories than used in \textcite{perrone20}. Moreover, we employ different network-based tools, enabling us to analyze different aspects of turbulent dispersion.

The transport network formalism is applied to the trajectories of tracer particles obtained from the direct numerical simulation of a fully developed turbulent flow in a channel at five different Reynolds numbers. To do so, we periodically release tracer particles inside the domain and track their motion over time. Using the trajectory data we build the network by measuring the exchange of fluid between fixed partitions of the wall-normal direction $y$ of the domain. These partitions, which are unbounded in the spanwise and streamwise directions and have constant wall-normal width, are the nodes of the network; the motion of particles generates the connections between the nodes.
Starting from the network, we describe the spatial inhomogeneities of vertical mixing, its dependency on the Reynolds number and the relationship between particle motion and the underlying turbulent flow phenomena. 

The paper is organized into five sections: section \ref{sec:meth} describes the methods employed, in particular the numerical simulations used to obtain the particle trajectories (\ref{sec:lagr_data}) and the procedure employed to build the transport network (\ref{sec:net_def}). Section \ref{sec:res} reports the results obtained from the network-based analysis. Finally, section \ref{sec:discu} contains the discussion of the results and section \ref{sec:concl} some final remarks.

\FloatBarrier
\section{Methods}
\label{sec:meth}
\subsection{Lagrangian data}
\label{sec:lagr_data}
The Lagrangian data used in this work were obtained by integrating trajectories of tracer particles inside a numerically simulated turbulent flow field. Five direct numerical simulations (DNSs) were performed, at friction Reynolds numbers $\ret = \delta u_{\tau}/\nu = 180,\,265,\,395,\,590$ and $950$, where $\delta$ is the channel half-height, $\nu$ is the kinematic viscosity and $u_{\tau} = \sqrt{\tau_w/\rho}$ is the friction velocity, with $\tau_w$ being the wall shear stress and $\rho$ the fluid mass density. The Navier-Stokes equations were solved with a pseudo-spectral method in a rectangular box of size $L_x\times L_y \times L_z$ \cite{kim_moin_moser_1987,canuto2012spectral,kuerten13}; periodic boundary conditions are applied along the streamwise $x$ and spanwise $z$ directions, while the no-slip condition is imposed at the walls ($y = 0$ and $y = 2\delta$). The main parameters used for the direct numerical simulations depend on the Reynolds number and are reported in table \ref{table:dns}.
The Navier-Stokes equations were made non-dimensional using outer-flow variables; accordingly, all the following results will be provided in the same units $x_i$ and $t$; these units are linked to the wall units by the relations $x_i = x_i^+ \nu/u_{\tau}$ and $t = t^+ \nu/u^2_{\tau}$.

\begin{table}
\centering
\caption{Main simulation parameters employed to generate the set trajectories at the five Reynolds numbers $\ret$ employed in this paper: domain size $L_x\times L_y \times L_z$, number of polynomials used for the simulation $N_x\times N_y \times N_z$ and time step of the trajectories $\Delta t$\label{table:dns}}
\begin{tabular}{cccc}
\toprule
$\ret$ & $L_x\times L_y \times L_z$ & $N_x\times N_y \times N_z$ & $\Delta t$\\
\midrule
180 & $4\pi\delta\times 2\delta \times 4\pi\delta/3$ & $384\times 193\times 192$ & $12.5\cdot 10^{-4}$\\
265 & $2\pi\delta\times 2\delta \times \pi\delta$ & $512\times 257\times 512$ & $10\cdot 10^{-4}$\\
395 & $2\pi\delta\times 2\delta \times \pi\delta$ & $512\times 257\times 512$ & $7.5\cdot 10^{-4}$\\
590 & $2\pi\delta\times 2\delta \times \pi\delta$ & $768\times 385\times 768$ & $6.25\cdot 10^{-4}$\\
950 & $2\pi\delta\times 2\delta \times \pi\delta$ & $768\times 385\times 768$ & $5\cdot 10^{-4}$\\
\bottomrule
\end{tabular}
\end{table}

Tracer particles are assumed massless so that their velocity $\mathbf{v}(\mathbf{x}_0,\,t)$ matches at any time that of the local, instantaneous Eulerian velocity field $\mathbf{v}(\mathbf{x}(\mathbf{x}_0,\,t),\,t)$, where $\mathbf{x}_0$ and $\mathbf{x}(\mathbf{x}_0,\,t)$ are the release location of a tracer and its position at time $t$, respectively. The trajectories are therefore integrated according to the ordinary differential equation $\text{d}\mathbf{x}/\text{d}t = \mathbf{v}(\mathbf{x}(\mathbf{x}_0,\,t),\,t)$ with the same second-order Runge-Kutta scheme as used in the explicit part of the integration of the Eulerian field. The time step used for the integration of trajectories depends on the Reynolds number and is reported in table \ref{table:dns}. Whenever a particle reaches one of the periodic boundaries, it is reinserted into the computational domain at the opposite side; particles cannot reach the solid walls, since they follow the fluid motion. The fluid velocity at the particle position is obtained by means of a trilinear interpolation of the velocity field; although higher order methods could be used, the use of a low order method does not affect the statistical accuracy of the set of trajectories, since the time step is sufficiently small \cite{kuerten13}.

Particles were released in a $N_l\times N_l$ grid located at $x = 0$; subsequent levels of the grid are spaced in the vertical direction $\Delta y = 2\delta/N_l$, while the spacing in the spanwise direction is $\Delta z = L_z/N_l$. $N_l$ is equal to $100$ for all Reynolds numbers, so that in any case $N_p = \num{10000}$ particles are released. Because of this layout, all particles are initially confined to one location in the streamwise direction of the channel and are therefore influenced by the instantaneous velocity field of that location. To evaluate the influence of the initial condition on the transient phase of the network evolution, we released $N_b = 61$ subsequent sets of particles, each one composed by the same $N_l\times N_l$ grid of tracers located at $x = 0$, for each Reynolds number. We separated the release of the batches by a time appropriate to the Reynolds number of the simulation, \textit{i.e.} a time larger than the Lagrangian integral time scale; by doing so, we ensured that the different batches are uncorrelated. We integrated all the trajectories for a duration of at least $T = 10$ in order to analyze the entire transient phase of the mixing process before the Taylor regime is reached; accordingly, each set of trajectories is composed of $N_t = T/\Delta t$ discrete time steps. Lagrangian statistics obtained from the sets of trajectories employed in this work are made available in an online repository \cite{lagstat}.

\subsection{Transport network}
\label{sec:net_def}
We aim to transform the continuous description of the Lagrangian dynamics of the flow provided by the particle trajectories into a discrete network representation. Since our main focus is mixing in the wall-normal direction, we divided the channel in $N_l = 100$ equal levels along the $y$ direction, each with height equal to $2\delta/N_l$; these slabs are unbounded in the $x$ and $z$ directions. 
We mapped each trajectory $\mathbf{x}(\mathbf{x}_0,\,t)$ into a sequence of traversed levels; doing this for the entire set of particles, we were able to transform the three-dimensional trajectories into a succession of discrete states $\mathbf{s} \in \mathbb{R}^{N_l}$, where $s_i$ is the concentration of tracers in level $i$. 
Then, we built the transition probability matrix $\mathbf{P}(t_0,\,\tau)\in\mathbb{R}^{N_l\times N_l}$. Each entry $P_{ij}$ of the transition matrix is defined as the probability that a tracer, located in level $i$ at $t = t_0$, ends up in level $j$ at $t = t_0+\tau$.  
We approximate this probability by the fraction of the particles released in level $i$ at time $t_0$ that is in level $j$ at $t = t_0+\tau$. The entries of the transition matrix are therefore computed as
\begin{equation}
\label{eq:transmat}
P(t_0,\,\tau)_{ij} = \frac{N_{i\to j} (t_0,\,\tau)}{N_i(t_0)},
\end{equation}
where $N_{i\to j} (t_0,\,\tau)$ is the number of particles, released in level $i$ at time $t_0$, that are located in level $j$ at time $t_0+\tau$; $N_i(t_0)$ is the total number of tracers released in level $i$ at the release time ($N_i(t_0) = 100$ for all cases).
Since particles cannot leave the domain, the rows of $\mathbf{P}(t_0,\,\tau)$ always add up to unity. On the other hand, this is not true for the columns of $\mathbf{P}(t_0,\,\tau)$; indeed, the sum of the $i$-th column of $\mathbf{P}(t_0,\,\tau)$ is the number of particles contained in level $i$ at time $t_o + \tau$ divided by $N_i(t_0)$. 

Formally, $\mathbf{P}(t_0,\,\tau)$ represents the transition process from the state $\mathbf{s}_{t_0}$ of the system at $t = t_0$ to the state $\mathbf{s}_{t_0+\tau}$ at $t = t_0+\tau$. 
The transition between subsequent states is determined solely by the transition matrix, so that $\mathbf{s}_{t_0+\tau} = \mathbf{P}^{\intercal}(t_0,\,\tau) \mathbf{s}_{t_0}$, where the superscript $^{\intercal}$ indicates matrix transposition.

The transition matrix $\mathbf{P}(t_0,\,\tau)$ can be interpreted in a straightforward manner as the weight matrix of a network \cite{newman2018networks}. A network $G(\mathcal{V}, \mathcal{E})$ is a structure which includes a set of nodes $\mathcal{V}$, representing discrete, interacting objects, and a set of links $\mathcal{E}$, which are the interactions between nodes. Networks may be directed, meaning that each link is an ordered pair of nodes and retains directional information, and weighted, \textit{i.e.} a value $w_{ij}\in\mathbb{R}$ is associated to each link $\left( i,\,j\right)\in\mathcal{E}$ to represent some relevant information. A natural representation of a weighted network is the weight matrix $\mathbf{W}$, defined as
\begin{equation}
\label{eq:weightmat}
W_{ij} = \left\lbrace\begin{matrix}
w_{ij}\text{ if } \left( i,\,j\right) \in \mathcal{E} \\
0\text{ if } \left( i,\,j\right) \notin \mathcal{E} .\\
\end{matrix} \right.
\end{equation}
Ignoring the weights of links leads to the definition of the adjacency matrix $\mathbf{A}$, whose entries are defined as 
\begin{equation}
\label{eq:adjmat}
A_{ij} = \left\lbrace\begin{matrix}
1\text{ if } \left( i,\,j\right) \in \mathcal{E} \\
0\text{ if } \left( i,\,j\right) \notin \mathcal{E} .\\
\end{matrix} \right.
\end{equation}

\begin{figure}
\centering
\includegraphics[width = \textwidth]{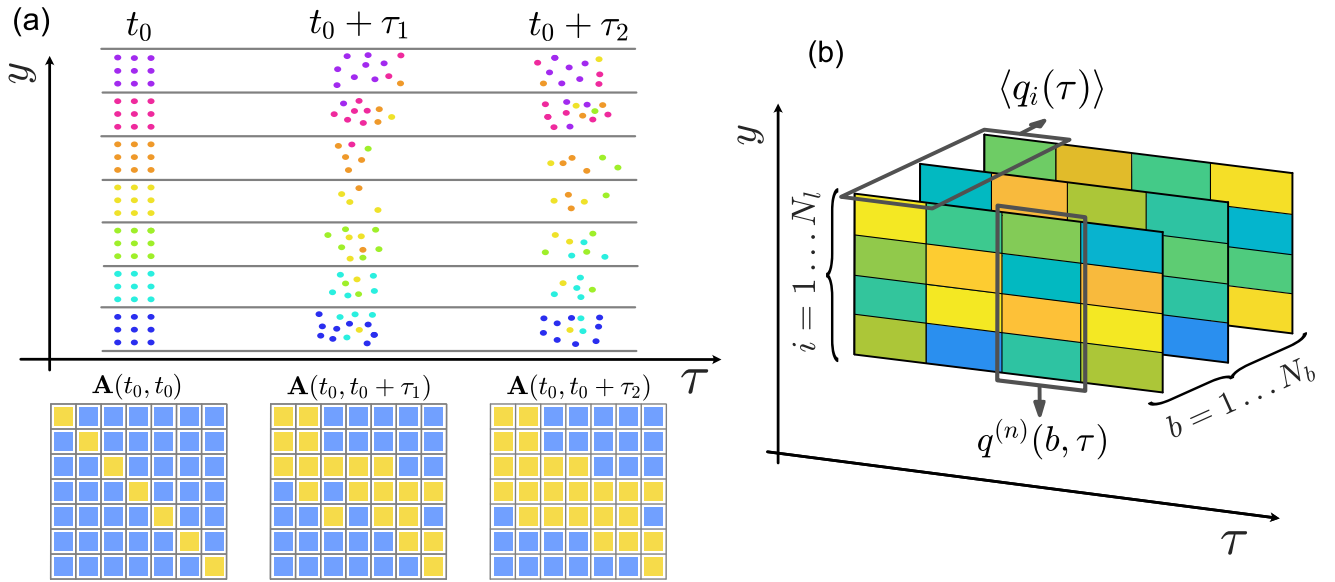}
\caption{(a) Schematic representation of the motion of tracers and the resulting network adjancency matrices: particles are marked with colors according to their release level and are tracked during the simulation; for each time considered, a network can be derived: the adjacency matrices of the networks generated from the particles in the example are shown in the lower panels (where a yellow square indicates an entry of $\mathbf{A}$ equal to 1). (b) Schematic of the different averages used in this work: a generic quantity $q_i(b,\tau)$, which is related to node $i$ and realization $b$, can be averaged over the height of the channel, $ q_i^{(n)}(b,\tau)$, or over different realizations at the same Reynolds number, $\langle q_i(\tau)\rangle$.\label{fig:1}}
\end{figure}

In this work, we set the network weight matrix $\mathbf{W}(t_0,\,\tau)$ to be equal to the transition matrix $\mathbf{P}(t_0,\,\tau)$. For each batch of particles released into the channel, we build a network that represents the transition between the state of particles at the moment of their release and the state after a time $\tau$, therefore quantifying the transfer of tracers between levels of the channel as a function of time. We obtain a temporal sequence of networks by keeping $t_0$ fixed and increasing $\tau$.  The networks resulting from the transition probability matrix are directed, because each link has a starting and ending node following the wall-normal direction of the motion of tracers in the interval $[t_0,\,t_0+\tau]$, and also weighted, because connections have an associated weight equal to the fraction of particles that move from one level to another. The networks are also time-dependent and spatially embedded, since their nodes have a definite (and stationary) location inside the physical domain.

The asymmetric and real-valued matrix $\mathbf{W}(t_0,\,\tau) = \mathbf{P}(t_0,\,\tau)$ fully describes the discretized motion of tracers and thus provides a simplified framework for the study of turbulent mixing. The scales of motion smaller than the level size are not explicitly displayed in the network representation, even if the integration of trajectories accounts for their effects. The sensitivity of the network to the number of levels $N_l$ will be further analyzed in the Appendix\ref{sec:sens}.
Furthermore, the adjacency matrix $\mathbf{A}(t_0,\,\tau)$ of the transport network may be constructed starting from its definition (\ref{eq:adjmat}). The entries $A_{ij}(t_0,\,\tau)$ of the adjacency matrix are equal to one if at least one tracer released in level $i$ at $t_0$ is located in level $j$ at $t_0+\tau$. Therefore, the adjacency matrix gives information about the material connection of two levels of the network at time $\tau$, neglecting the strength of the connection.
The procedure used to build the network is illustrated in figure \ref{fig:1}(a): at each time $\tau$, the position of tracers is binned into the $N_l$ discrete levels, while also keeping track of their release location. The lower panels of \ref{fig:1}(a) show the time-dependent adjacency matrices built from the motion of tracers.

Because of the dispersion process experienced by tracers, as the delay $\tau$ grows more and more particles leave the levels they were released in and drift away from them, so that new links of the network are activated. For sufficiently long delays $\tau$, networks built upon different sets of tracers become statistically similar independently of the flow field at the time of release, since the underlying dispersion process has reached the asymptotic regime identified by \textcite{taylor_diff}. This happens when tracers have lost all memory of their starting location and are homogeneously distributed across the channel, which occurs for times far larger than the Lagrangian integral timescales of the flow. At this point, links in the network are randomly distributed and their properties depend solely on the ratio $N_p/N_l^2$ (if the normalizing factor in equation \ref{eq:transmat} is ignored, the weights follow a Poisson distribution with mean equal to $N_p/N_l^2 = 1$ \cite{perrone20}).

Whenever each batch $b$ is released into the channel, we compute $N_t$ network weight matrices $\mathbf{W}(b,t_0,\,\tau)$ (being $\tau = j\Delta t$, with $j = 1\ldots N_t$). The network matrix depends only on the delay $\tau$ and on the properties of the set of trajectories $b$. In the following we will omit the argument $t_0$, which we always set to be the release instant of the tracers, so that $\mathbf{W}(b,\tau) = \mathbf{W}(b,t_0,\,\tau)$. By doing so, we obtain for each set of tracers a succession of weight matrices (and the corresponding adjacency matrices $\mathbf{A}(b,\tau)$), covering the entire evolution of the trajectories across the channel. 
Repeating this for all realizations gives us $N_b$ successions of networks at each Reynolds number, which are uncorrelated because they originate from particle sets whose trajectories are in turn uncorrelated. 

Since particles of subsequent realizations are released after a temporal spacing larger than the integral timescales of the flow, the uncorrelated velocity fields in which particles are inserted originate highly diverse successions of networks $\mathbf{W}^{(b)}(\tau)$. An ensemble analysis of the properties of the network at each Reynolds number can therefore be obtained by averaging the properties of the networks resulting from the $N_b$ different realizations. 
Releasing several, uncorrelated batches of tracers allowed us to study in great detail the effects of an irregular velocity field on the network transient by sampling the domain at different times; moreover, since particles are released in a single plane at $x = 0$, we are able to assess the effects of different realizations of an inhomogeneous velocity field. 
In the following, we mainly deal with quantities related to the nodes of the network, which are the spatial levels in which the channel is divided; these quantities are dependent on the physical location of the $i$-th node, on the realization $b$ and on the delay $\tau$. To provide statistical information about the behavior of the network, we define two different averages for the generic $i$-th node-related quantity $q_i(b,\tau)$, one along the wall-normal direction $ q^{(n)}(b,\tau) = \sum_{i = 1}^{N_l} q_i(b,\tau)/N_l$ and another between different realizations of the network $\langle q_i(b,\tau)\rangle = \sum_{b = 1}^{N_b} q_i(b,\tau)/N_b$. The $y$-average $\bullet^{(n)}$ and the ensemble average $\langle \bullet \rangle$ are illustrated in figure \ref{fig:1}(b).

\FloatBarrier
\section{Results}
\label{sec:res}
We focus our attention on the transient phase of the mixing process and its dependence on the Reynolds number, $\ret$, as captured by the transport network. The succession of networks resulting from the motion of tracers can then be used to extract information about the mixing process. We analyze the network, which is a reduced-order representation of particle motion, with different tools in order to characterize its main features and obtain results about the underlying physics. Section \ref{sec:degree} shows how the mixing process evolves from the network centrality perspective; section \ref{sec:cycles} provides a methodology to find the long-time effects of coherent motions inside the flow on particle dynamics using the network; section \ref{sec:eig} deals with the relation between the spectral properties of the network and the short-time properties of mixing. Finally, section \ref{sec:temp} gives some information about the role of the walls of the channel in relation to the temporal properties of the network.

\subsection{Degree centrality}
\label{sec:degree}

We aim to quantify the dispersion process of massless tracers in a turbulent channel flow, with a particular focus on the impact of the Reynolds number on its transient temporal evolution, on the role of wall-induced inhomogeneities and on the influence of the release condition. As time $\tau$ grows, particles move farther and farther from the starting point of their trajectory, as they are driven away by the mean flow advection and the dispersion induced by turbulent fluctuations of the velocity. In particular, mixing in the wall-normal direction, which is the focus of this work, is primarily influenced by the fluctuations of the $y$ component of the velocity, while the mean velocity normal to the wall is zero. 
We release large groups of particles at several wall-normal locations; as time grows, they will move away from their starting location and disperse across the $y$ direction of the channel. From the coarse point of view employed in this work, tracers leave the level they were released in and enter new different levels. Since turbulent channel flow is inhomogeneous in the $y$ direction, groups of tracers released from different levels will behave differently, according to the properties of the underlying velocity field.

\begin{figure}
\centering
\includegraphics[width = \textwidth]{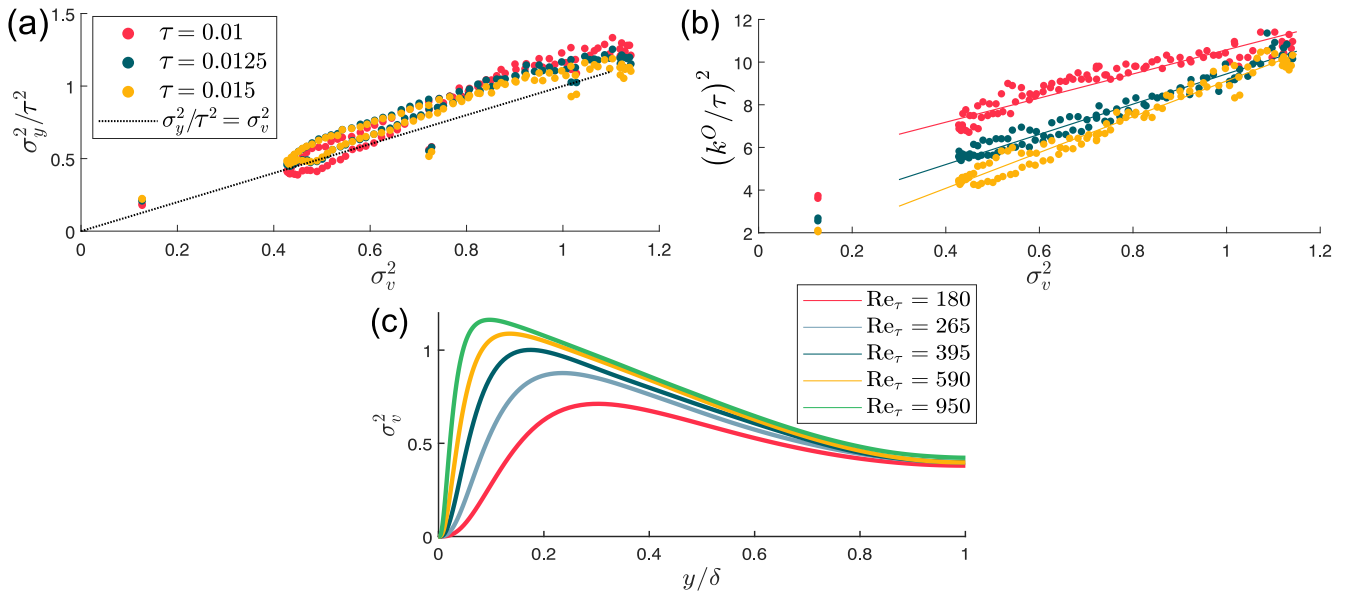}
\caption{(a) Variance of the position of particles released from each node and divided by $\tau^2$, plotted against the velocity variance at the release location, at $\ret = 950$. (b) Squared outgoing degree $\left(k^O\right)^2$, averaged over different realizations at $\ret = 950$ and divided by $\tau^2$, versus the velocity variance. (c) Wall-normal velocity variance profiles across the channel half-height.\label{fig:2}}
\end{figure}

For times much shorter than the Lagrangian velocity integral timescale $\mathcal{I}_L$, all particles that share the release location are still confined in the neighborhood of that location; moreover, their velocity is still correlated to the velocity at their release. Taylor's dispersion theory, adapted to inhomogeneous and anisotropic flows, prescribes that the dispersion of particles for $\tau\ll \mathcal{I}_L$ is ruled only by the variance of the velocity at the tracers' release location \cite{taylor_diff,stelzenmuller2017lagrangian}. In particular, the variance $\sigma_y^2(y_0,\tau)$ of the wall-normal position at time $\tau$ of particles released from the same wall-normal location $y_0$ grows ballistically as
\begin{equation}
\label{eq:variance2}
\sigma_y^2(y_0,\tau) = \tau^2\sigma_v^2(y_0,0) + \mathcal{O}(\tau^3),
\end{equation} 
where $\sigma_v^2(y_0,0) = \overline{v_y v_y}(y_0,0)$ is the wall-normal velocity variance (shown for all Reynolds numbers in figure \ref{fig:2}(c), with the overline indicating average over particles that share the same location $y_0$. Figure \ref{fig:2}(a) shows the validity of this relationship in the channel flow at $\ret = 950$ for three different times $\tau\ll \mathcal{I}_L$ by plotting the position variance divided by $\tau^2$ against the velocity variance for each of the $N_l$ release locations used in this work.

On the other hand, for times comparable or longer than the Lagrangian timescale, these assumptions become invalid because the velocity of tracers is no longer strongly correlated to that at their release. As particles move away from their starting level, they encounter regions with properties that are different due to the inhomogeneity of the flow.
Due to the advancing diffusion, the dispersing cloud of tracers samples a statistically inhomogeneous velocity field, so that a simple prediction of its evolution is no longer possible.
Moreover, the dispersion of tracers released near a wall becomes skewed by the presence of the boundary and particles can only diffuse in one direction. Also, clouds of tracers become entrained and influenced by coherent motions with a relatively large scale, especially in the near-wall region which is characterized by a range of coherent flow structures.

In the discrete approach proposed in this work, one can quantify the dispersion of a group of particles released from a channel level by measuring the number of other levels reached by these particles as a function of time. 
Conversely, mixing into the $i$-th level can be quantified by measuring the number of release levels whose particles are present in level $i$ at any time $\tau$. The number of levels whose particles have reached level $i$ at time $\tau$ is a measure for backward-in-time dispersion. 

From the transport network perspective, quantification of the number of levels encountered by dispersing tracers during their motion (forward or backward in time) is done by measuring the number of connections leaving and entering a node (which, by definition, corresponds to a level of the channel). The number of connections of a node $i$, regardless of their weight, is measured by the degree centrality $k_i$. For directed networks, both the outgoing $k_i^O$ and ingoing degree $k_i^I$ can be defined, accounting for the directionality of connections \cite{newman2018networks}. 
Both these measures can be computed, at any time $\tau$, as the sum of rows or columns respectively of the adjacency matrix $\mathbf{A}(\tau)$, so that $k_i^O(\tau) = \sum_{j=1}^{N_l} A_{ij}(\tau)$ and $k^I_i(\tau) = \sum_{j=1}^{N_l} A_{ji}(\tau)$. In the example shown in figure \ref{fig:1}(a), the outgoing degree of the nodes (from top to bottom) is $\mathbf{k}^O(\tau_1) = \lbrace 2,\,2,\,5,\,5,\,3,\,2,\,1\rbrace$ and $\mathbf{k}^O(\tau_2) = \lbrace 2,\,2,\,4,\,7,\,5,\,3,\,1\rbrace$, while the ingoing degree is $\mathbf{k}^I(\tau_1) = \lbrace 3,\,4,\,2,\,2,\,3,\,3,\,3\rbrace$ and $\mathbf{k}^I(\tau_2) = \lbrace 4,\,5,\,3,\,3,\,3,\,3,\,3\rbrace$.

Similarly to the short-time wall-normal position variance $\sigma_y^2(y_0,\tau)$, also the degree centrality for $\tau\ll T_{\mathcal{L}}$ is influenced only by the local velocity field. Figure \ref{fig:2}(b) shows how, for the same time instants considered in figure \ref{fig:2}(a), the squared outgoing degree $\left(k_i^O(\tau)\right)^2$ divided by $\tau^2$ is proportional to the wall-normal velocity variance $\sigma_v^2(y_0,0)$ measured at the center of the level (which is the release location of tracers). Unlike the wall-normal position variance $\sigma_y^2(y_0,\tau)$, the squared outgoing degree does not grow proportional to $\tau^2$. This happens because of the different sampling of the wall-normal size of the cloud of particles; while the position variance varies continuously, the degree changes discontinuously, and only when a tracer enters a previously unexplored level. The difference between the two sampling approaches is larger for shorter times $\tau$, when the particle distribution is narrower, and decreases afterwards.
Still, the degree is able to measure accurately the differences in the dispersion of particles due to the fluctuating velocity field at short times, because of its initial proportionality to the velocity variance.

\begin{figure}
\centering
\includegraphics[width = \textwidth]{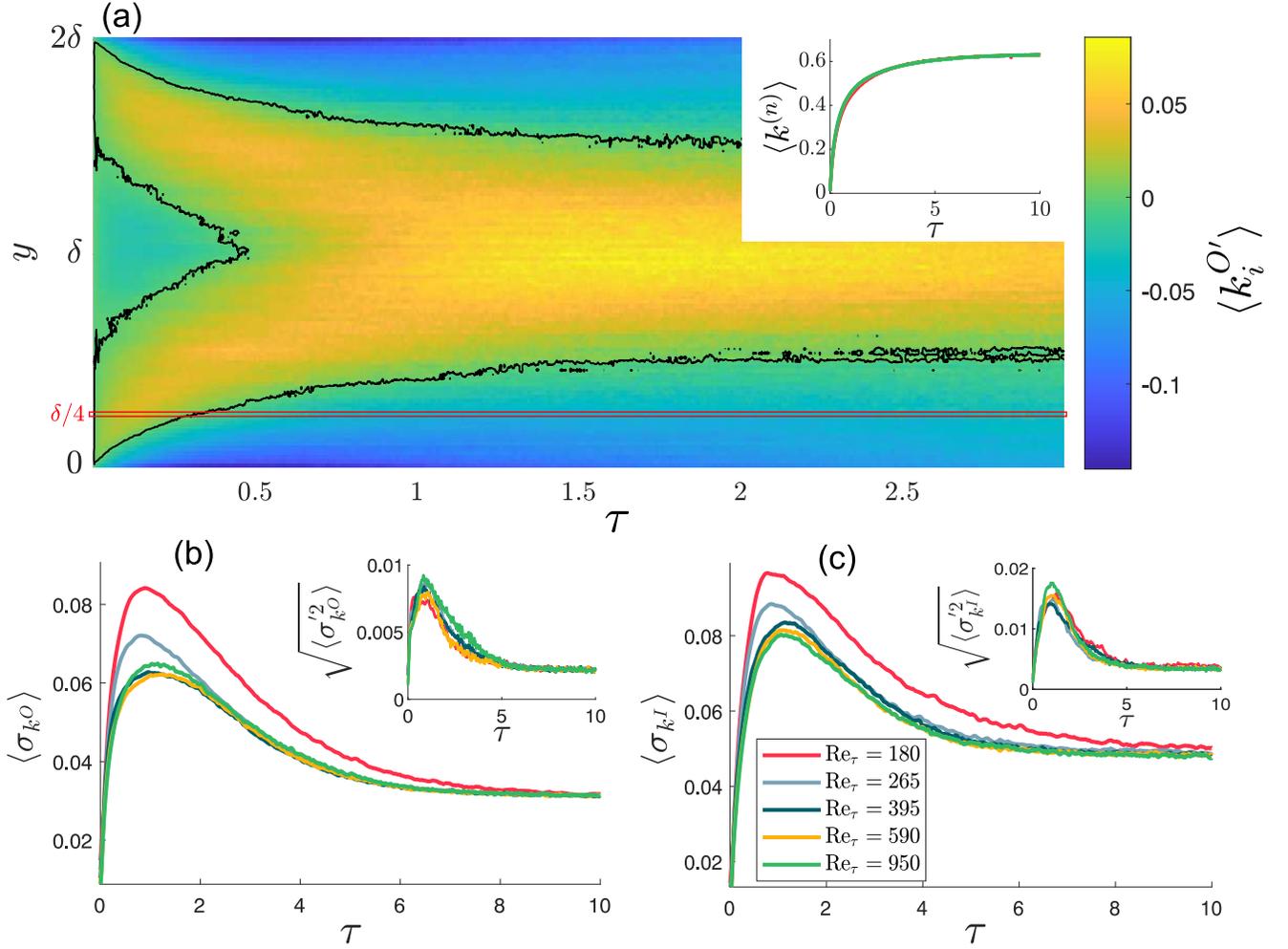}
\caption{(a) Fluctuation of the outgoing degree around its $y$-average, ensemble averaged at $\ret = 950$ as a function of $\tau$ (the black line marks the point for which $\langle k_i^{'}\rangle = 0$; also, the level located at $y = \delta/4$ is marked in red); the inset shows the $y$- and ensemble averaged degree for all five Reynolds numbers. (b)-(c) Ensemble averaged standard deviation $\langle\sigma_k\rangle$ of the outgoing and ingoing degree respectively, while the insets show the standard deviation across different realizations of $\sigma_k$, \textit{i.e.} $\langle \sigma_k^{'2}\rangle^{1/2} = \langle (\sigma_k - \langle \sigma_k \rangle)^{2}\rangle^{1/2}$.\label{fig:3}}
\end{figure}

For longer times, the degree centrality is able to quantify the extent of dispersion and mixing in each level of the channel, embedding the complexity arising from the inhomogeneous geometry of the flow. Due to this capability, both the spatial inhomogeneities and the differences between realizations of the dispersion of tracers can be studied by means of the evolution of the degree centrality.
As expected, the $y$- and ensemble-averaged degree $\langle k^{O,(n)}(\tau)\rangle = \langle k^{I,(n)}(\tau)\rangle$ (shown in the inset of figure \ref{fig:3}(a)) grows monotonically with $\tau$, since particles everywhere tend to disperse away from their starting level and enter new levels. 
The monotone growth of the mean degree reflects the transition from the ordered initial condition to the fully mixed, random final state of the system. After some time, the rate at which tracers enter new levels balances the rate at which they leave those levels, so that the averaged degree reaches a stationary asymptote; since at this point the weights $W_{ij}$ of the network are Poisson distributed, as stated in section \ref{sec:net_def}, also the asymptotic value of the degree can be analytically determined and is equal to the number of nonzero weights, \textit{i.e.} $\lim_{\tau\to\infty} \langle k^{(n)}\rangle = 1-1/\mathrm{e} \approx 0.63$ \cite{perrone20}. Also, the $y$-averaged ingoing degree $k^{I,(n)}(\tau)$ is equal to the $y$-averaged outgoing degree $k^{O,(n)}(\tau)$, because all connections leaving nodes of the network must enter other nodes, since tracers cannot leave the domain.

To show the dependence of the degree on the wall-normal coordinate, we calculate its spatial fluctuations around its $y$-averaged value, $k_i^{'}(b, \tau) = k_i(b,\tau) - k^{(n)}(b,\tau)$. Figure \ref{fig:3}(a) shows the fluctuation of the outgoing degree $\langle k_i^{O'}(\tau)\rangle$, ensemble-averaged at $\ret = 950$; the ingoing degree shows a similar appearance, as do the ensemble averages at the other Reynolds numbers. The fluctuation of the degree reveals the inhomogeneities of the dispersion process, which are determined by the effects of the walls. 
While overall the degree grows everywhere, it does so differently depending on the position: near the walls, the fluctuations of the degree are negative, indicating that there $\langle k_i(\tau)\rangle$ grows more slowly than its $y$-average and, consequently, tracers in that region disperse more slowly than in other parts of the channel. Moreover, the region in which $\langle k_i^{'}(\tau)\rangle<0$ expands over time proportionally to $\sqrt{\tau}$ for $\tau<1$; after this initial growth, the size of the region in which the fluctuations of the degree are negative occupies about one quarter of the channel height on each side until the transient phase ends. Instead, near the center of the channel and for $\tau<0.5$, the fluctuations $\langle k_i^{'}(\tau)\rangle$ of the degree are small, so that the ensemble-averaged degree grows as fast as its $y$-average. In the remaining regions of the channel, the degree grows faster than its average. 
In particular, there are two symmetric peaks, initially located very close to the walls, that join at the middle of the channel at $\tau \approx 0.5$.
Finally, for large values of $\tau$ ($\tau>6$, not shown in the figure), the fluctuations of the degree become rather small and, more importantly, uniformly distributed across the channel, indicating that the transient phase of the dispersion process has ended.

As shown by figure \ref{fig:3}(a), the degree does not only depend on the wall-normal position of the level inside the domain, but this dependency also varies with time. As an example, the ensemble averaged fluctuation of the degree $\langle k_i^{'}(\tau)\rangle$ of a level located at $y = \delta/4$ (marked in red in figure \ref{fig:3}(a)) is initially greater than zero, then decreases and becomes negative at about $\tau = 0.4$, and remains negative until the transient phase ends. This indicates that the diffusion in that location is initially stronger than the average; indeed at that location the variance of the wall-normal component of the velocity is the highest of the entire channel. However, after some time particles have moved away from there and have entered other levels.
In particular, particles that have moved towards the wall now diffuse more slowly than the average, thus causing a decrease of $\langle k_i^{'}(\tau)\rangle$ for the level considered. Conversely, particles that are released near the center of the domain after some time reach regions of the channel with stronger wall-normal velocity fluctuations (which are, ultimately, the main driver for wall-normal dispersion), so that the fluctuation of the degree for the centerline levels increases with time.

To effectively compare the fluctuations of the degree at different Reynolds number, we calculate their root mean square 
\begin{equation}
\label{eq:rmsk}
\sigma_k(b,\tau) = \sqrt{\frac{1}{N_l}\sum_{i = 1}^{N_l}\left(k_i^{'}(b,\tau)\right)^2},
\end{equation}
which quantifies the spatial inhomogeneity of mixing inside the channel. Figures \ref{fig:3}(b)-(c) show the ensemble averaged values of the root mean square of the degree, \textit{i.e.} $\langle \sigma_{k^O}(\tau)\rangle$ and $\langle \sigma_{k^I}(\tau)\rangle$, for the outgoing and ingoing degree at all Reynolds numbers used in this work. Independently of $\ret$, the standard deviations of the ingoing and outgoing degree grow rapidly, reach a peak at approximately $\tau = 1$ and then slowly decrease towards a stationary asymptotic value.

The standard deviations of the degree for $\ret\geqslant395$ are almost independent of the Reynolds number, and they collapse onto one single curve when outer units are used to normalize time. On the other hand, $\langle \sigma_k(\tau)\rangle$ at $\ret = 180$ and $\ret = 265$ is significantly higher during the entire transient phase for both the ingoing and outgoing degree. We confirmed the independency between realizations at $\ret = 180$ and 265 and those at higher Reynolds numbers with the Kruskal-Wallis test, while at the same time we found that realizations at $\ret\geqslant 395$ are not independent \cite{kruskalwallis}.  We analyzed the relationship between the collapse of the standard deviation and the discretization employed to build the network in the Appendix\ref{sec:sens}.
The main cause for the increased inhomogeneity of the degree at $\ret = 180$ is to be found in the near wall region, where the fluctuations of the degree are larger (in absolute value) than at the other Reynolds numbers. 

While the ingoing and outgoing degree mostly show a similar behavior, we note that the spatial standard deviation of the ingoing degree is in general higher than that of the outgoing degree. This means that the mixing of tracers into the levels of the channel takes place more heterogeneously than the dispersion of particles leaving these levels. 
The larger standard deviation of the ingoing degree with respect to the outgoing degree is caused by the fact that the ingoing degree is higher than the outgoing degree near the center of the channel, while it is lower towards the boundaries. Since the ingoing and outgoing degree centralities are connected to dispersion of particles backward and forward in time, the differences between $\langle \sigma_{k^I}\rangle$ and $\langle \sigma_{k^O}\rangle$ signify a temporal asymmetry in the dispersion process.
Recent research has linked the presence of temporal asymmetries in particle dispersion to the irreversibility of turbulence and to the dissipative flux of energy from large to small flow scales \cite{jucha14, bragg, polanco18}.
A relation for the difference between the backward and forward in time variance of particles' position, which share some similarities with the ingoing and outgoing degree as seen at the beginning of this section, can be derived for short times $\tau$. In particular, including terms of order $\mathcal{O}(\tau^3)$ in equation \ref{eq:variance2} yields $\sigma_y^2(y_0,\tau) = \tau^2\sigma_v^2(y_0,0) + \overline{v_y a_y}(y_0,0)\tau^3 + \mathcal{O}(\tau^4)$, where $v_y$ and $a_y$ are the wall-normal components of particle velocity and acceleration; the difference between backward and forward dispersion is therefore
\begin{equation}
\label{eq:disp_diff}
\sigma_y^2(y_0,-\tau) - \sigma_y^2(y_0,\tau) = -2\overline{v_y a_y}(y_0,0)\tau^3 + \mathcal{O}(\tau^5),
\end{equation}
which quantifies the temporal asymmetry of the dispersion of a cloud of tracers released at a wall-normal coordinate $y_0$ \cite{polanco_thesis}. We measured the value of $-2\overline{v_y a_y}(y_0,0)$ and found that it is positive near the center of the channel and negative at distances from the wall lesser than $\delta/2$ (except for the near-wall region, where particles moving towards the wall have to decelerate and thus $-2\overline{v_y a_y}(y_0,0) > 0$); this agrees with the difference between the ingoing and outgoing degree, which at short times is negative at wall distances $y\lesssim\delta/2$ and positive elsewhere.

Finally, the insets of figures \ref{fig:3}(b) and \ref{fig:3}(c) show the standard deviation across realizations of $\sigma_k$, defined as $\sqrt{\langle \sigma_k^{'2}\rangle} = \sqrt{\langle (\sigma_k - \langle \sigma_k \rangle)^{2}\rangle}$, both for the outgoing and ingoing degree. This quantity represents the variability due to the differences between different realizations of the spatial inhomogeneity of the diffusion process, as reported by the degree centrality. The Reynolds number does not seem to play a significant role here, meaning that the intra-Reynolds variability is independent of $\ret$. In particular, the differences between transport networks $\mathbf{W}(b,\tau)$ originating from different sets $b = 1\ldots N_b$ of trajectories at the same Reynolds number, which are due to the differences between the velocity fields in which subsequent batches of particles are released, do not depend on $\ret$.

\subsection{Cycles}
\label{sec:cycles}

\begin{figure}
\centering
\includegraphics[width = \textwidth]{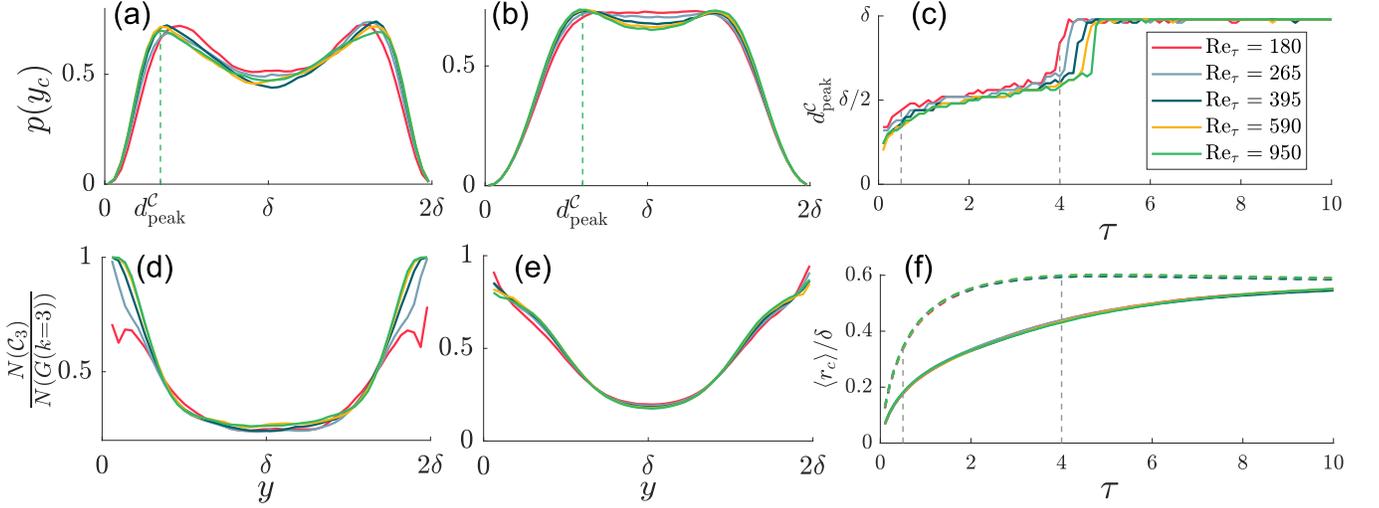}
\caption{(a)-(b) Concentration of cycles' centroids in the transport networks $\mathbf{W}(\tau = 0.5)$ and $\mathbf{W}(\tau = 4)$, ensemble averaged at five different Reynolds numbers. (c) Distance of the peaks of the concentration of cycles from the wall, as a function of time. (d)-e) Prevalence of cycles in the same networks as the previous panels, computed as the ratio between the number of cycles and the number of all subgraphs with three nodes. (f) Mean radius of cycles (solid lines) and of subgraphs that are not cycles (dashed lines), as a function of time; the vertical lines show the two time instants of figures \ref{fig:4}(a) and (b).\label{fig:4}}
\end{figure}

As tracers spread past the immediate vicinity of their release location, the inhomogeneities present in the fluctuating turbulent velocity field influence their dispersion and introduce complex effects which are captured by the degree centrality of the transport network representation. 
Moreover, in addition to the statistical properties of the velocity field, the dispersion of tracers is also influenced by the organized motions present in wall-bounded flows at a large range of scales, which are present at all Reynolds numbers. In particular, particles located near the wall become entrained in quasi-streamwise and hairpin vortices, which populate the logarithmic and outer regions of the flow; these structures determine the presence of intense wall-normal motions that eject low speed fluid (and the tracer particles therein) into the outer flow and the other way around \cite{smits}.
Thus, tracers sample the underlying velocity field; the more they are spread across the channel, the larger the flow structures they sample and interact with are. 

Small subsets of tracers that are close in space may become entrained in simple coherent structures, and in turn inherit their common and recognizable features. This translates, in the spatially coarse perspective employed in this work, into simple patterns of fluid exchange between levels of the channel. Tracers entrained in a wall-normal cyclic motion will move between a subset of the levels of the channel in a loop, at least for the time in which the underlying flow structure remains coherent.

As an example, a looping pattern involving three nodes $i$, $j$ and $k$ inside the transport network (formally, a cyclic subgraph of $\mathbf{W}(\tau)$), is generated when some tracers released from level $i$ reach level $j$ at time $\tau$, while at the same time tracers leaving $j$ and $k$ reach levels $k$ and $i$, respectively. In the transport network framework, the analysis of cyclic subgraphs has been used to uncover periodic trajectories in steady or periodic flows and cyclic motions in unsteady flows \cite{rodriguez}. In general, patterns contained in a complex network that occur significantly more than what would happen in a random graph are called \textit{motifs}. The study of motifs allows us to detect relevant processes taking place on the network and contributes in determining the complex behavior of the dynamical system represented by the network \cite{schwarze2020motifs,alon2007network}.

In order to identify motifs on the transport network $\mathbf{W}(\tau)$, we employed the algorithm of \textcite{wernicke}, which performs a search of the network for subgraphs of a given number of nodes and given structure. In particular, we searched the network for cyclic subgraphs with three nodes, \textit{i.e.} a subset of three nodes of the network that are connected in a directed loop. Structures comprising three nodes are of particular interest since they are linked to the definition of the clustering coefficient \cite{fagiolo}. Additionally, cyclic subgraphs composed of a larger number of nodes in the transport network identify other periodic structures with longer periods \cite{rodriguez}.

To determine whether the cyclic subgraphs found with the aforementioned procedure are indeed motifs, \textit{i.e.} they contribute significantly to the behavior of the network, we analyzed their statistical relevance using as a null model a random network.
In particular, we computed, using the same breadth-first search algorithm as before, cycles of three nodes in a random network in which nodes have the same number of connections of the nodes of the real transport network $\mathbf{W}(\tau)$. We found that in the real transport network cycles occur more frequently and are also distributed differently across the height of the channel. Specifically, cycles in the random network are more frequent near the center of the channel than near the walls, while this is not the case for the real network.

Figures \ref{fig:4}(a) and \ref{fig:4}(b) show the spatial distribution of the centroids of cyclic subgraphs in $\mathbf{W}(\tau = 0.5)$ and $\mathbf{W}(\tau = 4)$, respectively; the results were ensemble-averaged between realizations at the same Reynolds number. We computed the centroid coordinate $y_c$ by averaging the spatial coordinates $y_i$ of the three network levels involved in each cyclic subgraphs. The concentration of cycles has a peak near $y = \delta/2$ and decreases to zero close to the wall, because of the finite size of cycles. With increasing time, the peaks of the distribution move towards $y = \delta$, eventually joining abruptly in the middle of the channel at about $\tau = 4$. The shift of the peak location in time occurs almost independently of the Reynolds number, although at $\ret = 180$ and 265 peaks are somewhat closer to the center of the channel and they join in the center slightly earlier.
The distance of the peaks from the wall $d_{peak}^c$ as a function of time $\tau$ is shown in figure \ref{fig:4}(c).
At the start of the simulation, cycles are more likely to be found near the solid boundaries, where the action of near-wall vortices impresses a swirling motion on particles which may result in the formation of cyclic patterns of links.

As the transport network $\mathbf{W}(\tau)$ contains information about the trajectories of particles between $t=0$ and $t=\tau$, then also the cycles found inside the network take into account the entire previous motion of tracers. As such, the increased presence of cycles is the result of the consecutive action of several vortical flow structures located in the near-wall region. Because of how the transport network is built, one needs only temporally sparse position data of particles to explore the existence and effects of swirling motions of tracers, thus easing the computational requirements and the need for additional data (\textit{i.e.} particle velocities and accelerations or vorticity fields).

Figures \ref{fig:4}(d) and \ref{fig:4}(e) show the ratio between the number of cyclic subgraphs and the number of all subgraphs with three nodes, across the height of the channel and at the same times $\tau$ as used for figures \ref{fig:4}(a) and \ref{fig:4}(b). Near the walls, cyclic subgraphs are prevalent over all other subgraphs, while their relative concentration diminishes away from the walls. Thus, cyclic motions of particles are dominant near the walls of the channel. This happens at all Reynolds numbers, even though at the lowest one ($\ret = 180$) the prevalence of cycles is not as high as in the other cases.

In figure \ref{fig:4}(f) we show the mean radius of cycles $r_c$, computed as the average distance between each node involved in the cycle and its centroid $r_c = \frac{1}{3}\sum_{i=1}^3 \lvert y_i-y_c\rvert$. The mean radius of cycles (solid lines) grows more slowly than the mean radius of subgraphs with three nodes that are not cycles (dashed lines). This happens independently of the Reynolds number, even at the lower Reynolds numbers.
The slower growth of the size of cycles implies that the diffusion of clouds of particles entrained in cyclic motions is impeded, since their size in average grows more slowly than that of clouds of tracers that do not take part in cyclic motions. While the effectiveness of single cycles in retaining tracers is unchanged across different Reynolds numbers, the stronger prevalence of cycles at the higher Reynolds numbers indicates a stronger inhibition of dispersion caused by vortical flow structures.

\subsection{Eigenvector centrality}
\label{sec:eig}

\begin{figure}
\centering
\includegraphics[width = \textwidth]{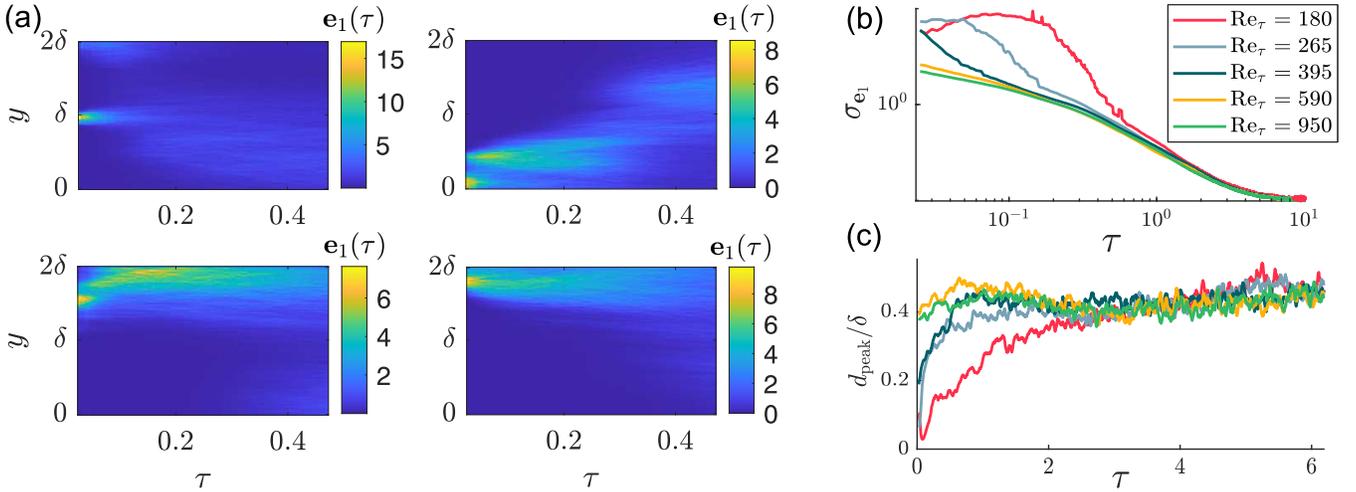}
\caption{(a) Eigenvector centrality $\mathbf{e}_{1}(\tau)$ for four different realizations of the transport network evolution at $\ret = 950$. (b) Temporal behavior of the standard deviation of the node eigencentrality, averaged over realizations at the five Reynolds numbers considered. (c) Distance between the node with the highest eigenvector centrality and the nearest wall, averaged over realizations at the five Reynolds numbers. \label{fig:5}}
\end{figure}

The transport network $\mathbf{W}(\tau)$ holds the discrete representation of particle mixing in the interval $\left[t_0,\,t_0+\tau\right]$. The properties of $\mathbf{W}(\tau)$ are in turn the result of the effects of the flow field, which is integrated over the same time interval $\left[t_0,\,t_0+\tau\right]$ to obtain the trajectories of tracers.
To evaluate the cumulative effects of the flow field, we analyze the eigenvector $\mathbf{e}_1$ associated to the leading eigenvalue $\lambda_1$ of $\mathbf{W}^{\intercal}(\tau)$.
The eigenvector $\mathbf{e}_1$ obtained from the transport network weight matrix represents the concentration along $y$ that particles would attain if left indefinitely under the influence of the flow as represented by the network $\mathbf{W}(\tau)$ at a single time $\tau$.
In particular, the eigenvector $\mathbf{e}_1$ is the state of the system (\textit{i.e.} the concentration of tracers) attained after infinite applications of the same process $\mathbf{W}(\tau)$ at a fixed time $\tau$, which is unique and independent of the initial state \cite{chaos}. 

Analyzing $\mathbf{e}_1$ allows us to isolate the effects of the flow in the time interval $\left[t_0,\,t_0+\tau\right]$ and capture the trend of particle motion. 
The nodes in the transport network where the local value of the state $e_{i,1}$ is higher tend to attract tracers, even if the concentration given by $\mathbf{e}_1$ cannot be achieved, due to the time dependency of the weight matrix $\mathbf{W}(\tau)$.
The analysis of the eigenvector $\mathbf{e}_1$ of networks $\mathbf{W}(\tau)$ as a function of time yields information about the overall motion of particles and the presence or absence of specific levels in the channel that act as attractors for particles at different times $\tau$.

In general, the eigenvector associated to the leading eigenvalue of a network weight matrix is a centrality measure, the \textit{eigenvector centrality}, which quantifies the influence of a node on the process taking place on the network. A node has a high eigenvector centrality if it is in turn connected to other nodes with high eigenvector centrality \cite{newman2018networks, bonacich}. As an example related to a traditional application of network theory, airport hubs usually have a high eigenvector centrality in the aerial transportation network, not because of the number of airports they are connected to (which is measured by the degree centrality), but because of the importance of connected airports and because of their influence on the entire network.
The spectral properties of $\mathbf{W}(\tau)$ have also been exploited to find sets of tracers that remain coherent in time, that is states that are not perturbed by the action of the flow, similarly to what happens for the limit state $\mathbf{e}_{1}$ \cite{froyland2007prl, froyland09}.

The instantaneous flow field in which particles are released determines their initial velocity $\mathbf{v}(\mathbf{x}_0,0)$ and, thus, the initial evolution of their trajectories, at least for times shorter than the Lagrangian integral timescale.
Since different sets of particles are released inside the channel into uncorrelated velocity fields, also the resulting sets of trajectories evolve differently in their initial phase, with a strong dependence on the initial condition. Consequently, also the different realizations of the transport network $\mathbf{W}(\tau)$ are highly diverse, even at the same Reynolds number. The trend of motion imposed on tracers can be analyzed by means of the study of the eigenvector centrality and its temporal evolution.

Figure \ref{fig:5}(a) shows the eigenvector centrality $\mathbf{e}_{1}(\tau)$ of the nodes for four different realizations of the network at $\ret = 950$, as representative of the eigenvector centrality temporal behaviour: a clear peak, with a variable location depending on the realization, occurs at a time $\tau$ comparable with the integral timescale. A similar behavior is found in all realizations.
Nodes with a high eigenvector centrality act as attractors for the motion of tracers, and the higher the eigenvector centrality, the higher the strength at which tracers are driven towards these nodes. The highly inhomogeneous velocity field experienced by tracers at their release tends to displace them towards such nodes. As time grows, the effects of the inhomogeneities are smoothed out by the dispersion of tracers and also the eigenvector centrality become more uniform across all nodes. In the end, the eigenvector centrality becomes uniform across the channel height; indeed, the concentration of tracers becomes uniform in the wall-normal direction in the long run due to continuity and the well-mixed condition, and so does the limit state \cite{sawford_wm}.
To briefly summarize the evolution of the limit state for different Reynolds numbers, we computed the spatial standard deviation of the eigenvector centrality of each node and then we ensemble averaged its value between realizations of the network at the same $\ret$; results are shown in figure \ref{fig:5}(b). 
The relatively high spatial standard deviation of the eigenvector centrality for short times highlights a non-uniform attraction of tracers across the channel height. Higher values of the standard deviation, as those found for the channel flow at $\ret = 180$, indicate a stronger imbalance between attracting and non-attracting nodes. This in turn shows that tracers at the lowest Reynolds number are more susceptible to attracting structures in the flow, probably because of the reduced turbulent dispersion.
After some time, the eigenvector centrality tends to spatial uniformity, so that its standard deviation quickly decreases. 

Additionally, we show in figure \ref{fig:5}(c) the ensemble-averaged distance of the eigenvector centrality peak from the nearest wall of the channel, \textit{i.e.} the position of the nodes that act as attractors during the very first phase of the dispersion of tracers. The eigenvector centrality peaks of different realizations at $\ret = 590$ and $\ret = 950$ are almost always uniformly distributed over the interval $[0,\,\delta]$ (albeit slightly skewed towards the walls), thus their ensemble-averaged distance from the wall is approximately $\delta/2$; the same happens, after a short transient, for $\ret = 265$ and 395. Instead, at $\ret = 180$, peaks are located much closer to the walls.
Thus, near wall levels of the channel act as sinks for tracers, at least for some time (the eigenvector centrality becomes almost uniformly distributed after $\tau \approx 2$). This is also the cause for the increased standard deviation found in figure \ref{fig:5}(b).

\FloatBarrier
\subsection{Temporal properties of links}
\label{sec:temp}

\begin{figure}
\centering
\includegraphics[width = \textwidth]{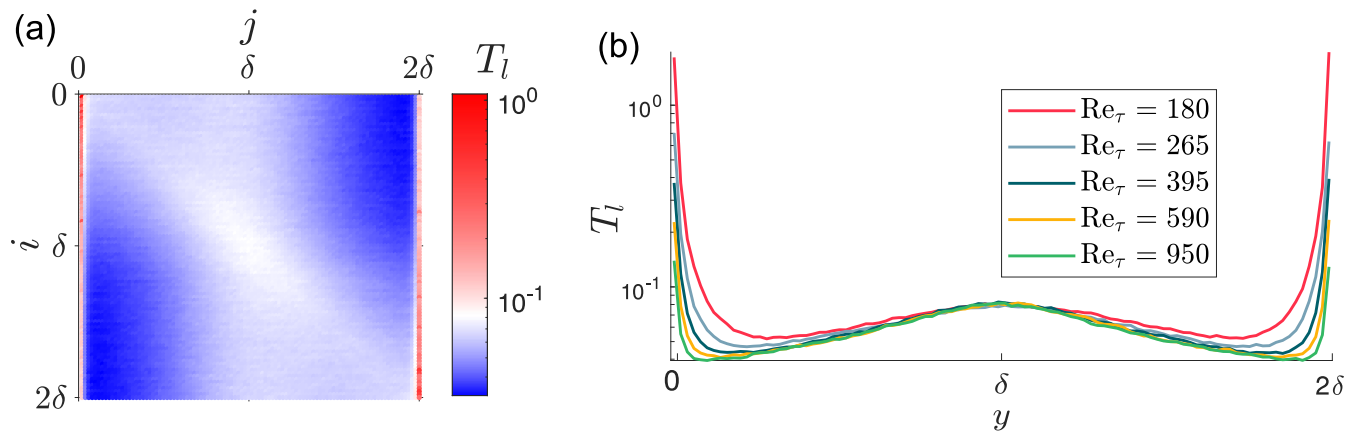}
\caption{(a) Mean link duration for each pair of nodes $\left( i,\,j\right)$, averaged over realizations at $\ret = 950$ (in the labels, the $y$ coordinate of nodes is shown). (b) The same quantity shown in detail for links starting from the centerline level, averaged over realizations at all Reynolds numbers\label{fig:6}}
\end{figure}

The peculiar effect of the walls on turbulent mixing is also evident from the analysis of the duration of links, \textit{i.e.} the time that a particle moving from level $i$ to level $j$ of the channel (and thus activating a link $\left(i,\,j\right)$) spends in level $j$ and maintains the link active. The temporal duration of links can be measured as the time a given link $(i,\,j)$ is continuously active in the evolving network $\mathbf{W}(\tau)$. In particular, we focus on the mean duration of links $T_l$, which we compute for each pair $(i,\,j)$ of nodes as the average temporal duration of the links connecting these two levels. $T_l$ is shown in figure \ref{fig:6}(a) for the channel flow at $\ret = 950$, where each entry $(i,\,j)$ of the matrix displays the value of the mean duration of links starting in level $i$ and ending in level $j$; we ensemble-averaged these results over realizations of the network at $\ret = 950$. The resulting matrix-like representation is not symmetric, since it follows from a directed network with an asymmetric weight matrix.

It can be noticed that the mean duration of links depends mostly on the spatial location of the ending node $j$. Indeed, links ending in the levels closest to the wall have the longest permanence time (even an order of magnitude larger than in other regions of the channel). Particles that move close to the walls become trapped in these regions and spend a long time in the same level of the network, thus resulting in long duration links. 
This region of increased temporal duration encompasses the level of the channel nearest to the wall (which has a height equal to $2\delta/N_l$). Typical values of $T_l$ in this region range from $T_l\approx 0.1$ at $\ret = 950$ up to $T_l\approx 1$ at $\ret = 180$, which is one tenth of the entire simulation time.

The mean duration of links is also slightly larger than the minimum values of $T_l$ for links ending near the centerline and for links originating between levels close in space; indeed, values of $T_l$ near the diagonal of figure \ref{fig:6}(a) are larger than for other couples of connected levels.
As an example, links starting and ending near the center of the channel have a mean duration $T_l\approx 0.1$, independently of the Reynolds number; this is enough for a tracer moving with the bulk velocity to move a distance $\Delta x\approx 2.2\delta$ downstream at $\ret = 950$.
The increased duration of links connecting nearby levels shows how $T_l$ depends not only on the ending level of the link, but also on the starting level (albeit in a lesser way). The lasting effect imparted by the starting level on the duration of links suggests the existence of some sort of memory for tracers that move between levels close in space; indeed, the duration of links is prolonged if tracers came from a nearby level of the channel. Of course, such memory effects, which are related to the persistent correlation of the velocity field, are possible only for levels close in space.

Figure \ref{fig:6}(b) shows the mean duration of links starting from the level located at the centerline $y = \delta$, with respect to the ending node $j$ of the link, for all Reynolds numbers. The plots of figure \ref{fig:6}(b) correspond to a single row of the matrix shown in figure \ref{fig:6}(a).
The duration $T_l$ of links ending in levels near the center of the channel is the same at all Reynolds numbers considered.
Instead, $T_l$ is far higher for connections ending near the wall at all Reynolds numbers, as already noted previously; this increased duration indeed contributes to the lesser diffusive properties of the near-wall region, since it implies that particles located near the walls become trapped there.  
Moreover, the duration of links ending in the near-wall levels is higher for lower Reynolds number. If wall units are used, the link duration $T_l^+ = T_l u_{\tau}^2/\nu$ for the level nearest to the wall is nearly constant at the three higher Reynolds numbers, while it is almost three times larger at $\ret = 180$. 

\FloatBarrier
\section{Discussion}
\label{sec:discu}

Using the different network features analyzed in the previous sections, we are able to provide a multifaceted perspective on the dispersion process. While some features of dispersion appear in the network at all Reynolds numbers, we found that some properties are exclusive to higher or lower Reynolds number flows. Most notably, while dispersion of tracers near the walls is reduced in any case, different mechanisms come into play depending on the Reynolds number. Moreover, through the network we found that above $\ret = 395$ many network properties exhibit Reynolds number independence if scaled with outer units (with the spatial resolution $N_l = 100$). 

The number of connections established by a node via particle motion, \textit{i.e.} its degree centrality, is a concise measure of the local dispersion of particles in that node. Owing to the non-homogeneous nature of channel flow, we found the growth of the degree to be highly dependent on both the spatial location of nodes and the time $\tau$ from the release of particles.
At very short times $\tau\ll\mathcal{I}_L$, the degree centrality of each node grows proportionally to the wall-normal velocity variance in that node. Indeed, when groups of particles are still confined to a small spatial region, their dispersion is solely dominated by the local velocity fluctuations. As such, the initial evolution of the degree (figure \ref{fig:3}(a)) shows how initially dispersion grows faster in the near-wall region, coherently with the $y$ location of the wall-normal velocity fluctuations peak.
Differently, for times comparable or longer than the Lagrangian integral timescales of the flow, the wall-normal velocity variance cannot alone explain the observed dispersion of particles. We observe that the dispersion becomes stronger near the centerline of the channel and weaker near the walls; therefore it appears that phenomena other than the wall-normal velocity fluctuations come into play, resulting in an inhibition of near-wall dispersion. 
Moreover, at short (but comparable with $\mathcal{I}_L$) times, the high values of the eigenvector centrality in a few close nodes of the network shows the presence of a localized attractor, \textit{i.e.} a zone in the channel that transiently attracts tracers. This attractor influences the dynamics of the entire set of particles, even those that are distant from the high eigenvector centrality nodes. Therefore, the attractor characterizes the shift from a regime in which only local velocity fluctuations determine the evolution of dispersion to a stage when the entire range of scales of motion in the channel flow affects the dispersion of groups of particles.

The shift of peak dispersion from the near-wall region of the channel to the centerline is also associated with the time at which the spatial non-homogeneity of dispersion is the highest, as is shown by the peak at $\tau\approx 1.5$ of the degree standard deviation $\sigma_k$. Only after a rather long (compared to the integral timescale) transient, the inhomogeneities of the degree are smoothed out, as wall-normal mixing is complete and dispersion has entered an asymptotic phase similar to Taylor's regime in homogeneous flows.

The network analysis enables us to pinpoint the causes for the inhibition of near-wall dispersion: the very low intensity of near wall velocity fluctuations and the presence of cyclic motions of particles, each one with its varying relative importance at different Reynolds numbers. 
First, wall-normal velocity fluctuations near the wall have a reduced intensity with respect to the rest of the channel, so that particles that move towards the wall are less prone to being vertically displaced by turbulent velocity fluctuations. This is also evident, in the network perspective, from the increased duration of links ending near the wall, which is linked to the residence time of particles in those levels. 
The second mechanism, highlighted by the analysis of motifs, is linked to the presence of cyclic patterns of motion of particles between channel levels. 
Tracers that move persistently in cyclic paths are effectively prevented from reaching other levels since, even if there is an exchange of tracers as in the other regions of the channel, this happens between nodes interconnected in loop-like paths that do not promote the diffusion of tracers towards other nodes. Thus, the significant increase in the concentration of cycles found towards the wall is an obstacle to particle dispersion.
Cyclic motion, which results from the action of vortical flow structures hosted in the near wall region, is highly important near the walls; additionally, the slower growth of the radius $r_c$ of cyclic motifs than other non-cyclic structures indicates that the embedding of particles in such motion patterns contributes to the inhibition of their dispersion for an extended period of time.

While at the higher Reynolds numbers analyzed in this work ($\ret = 395,\,590,$ and $950$) cyclic motifs are highly prevalent near the walls and constitute nearly the entirety of the network structure, we found the cycles at lower Reynolds numbers to be less frequent, both absolutely and with respect to the totality of link patterns. 
Even if cycles have a role in slowing down dispersion, the reduction of the prevalence of cycles is not associated to an increase of dispersion at the lower Reynolds numbers, where instead dispersion is even more inhibited.
The slower near-wall dispersion results in an increased heterogeneity of the degree at $\ret = 180$ and 265 and greatly increases the temporal duration of links ending in the near-wall region, indicating prolonged residence of tracers in those regions. 
Also, attractors in the channel at lower Reynolds numbers are predominantly located near the walls (see figure \ref{fig:4}), indicating that the near wall levels greatly influence the overall dynamics inside the channel and that particles entering these levels tend to remain there for prolonged times.

At lower $\ret$ the main cause for the inhibited dispersion seems to be the reduced wall-normal velocity fluctuations near the walls, which result in longer link durations and increased presence of attractors in that region. Cyclic paths, instead, are less prevalent and their peak concentration is located farther from the wall. As the Reynolds number grows larger, the near-wall effects become confined to a smaller portion of the channel and cyclic structures acquire a dominant role in slowing down dispersion of tracers.
It is also possible that a larger prevalence of near-wall vortices such as that found at higher Reynolds numbers, while being detrimental to dispersion across long timescales, is beneficial in lifting particles away from the near wall region.

At the higher Reynolds numbers ($\ret = 395,\,590,$ and $950$) we found that many network properties are independent of the Reynolds number, if outer flow variables are used to normalize the time $\tau$. The reason is that the mixing process involves ever growing scales of motion inside the flow, as the clouds of tracers released inside each level grow larger. Additionally, the discretization employed to create the network divides the channel into levels whose height is independent of $\ret$ when scaled in outer units (since it is always equal to $2\delta/N_l$). The discretization effectively filters out from the network representation scales smaller than $2\delta/N_l$. The importance of the filtered scales is analyzed in the Appendix\ref{sec:sens}.
Accordingly, the network evolves independently of the Reynolds number as outer flow scales are dominant in the overall mixing process and in its network representation. With $N_l>100$, network properties remain network independent for $\ret\geqslant590$ and become slightly different at $\ret = 395$, indicating that the effects of the smaller scales are only important at lower $\ret$.
The outer-scale collapse happens to both global properties of the network, such as the degree $y$-average and standard deviation, and to local properties like the spatial concentration of cyclic motifs, with the notable exception of the temporal duration of links which scales with inner variables in the near-wall region. 
Indeed, local properties of network nodes, even those located inside the near wall region, are influenced by the links entertained by these nodes. For short times these links only connect nodes which are very close in space, and thus node properties reflect the local flow features (\textit{i.e.} the velocity fluctuations as shown in figure \ref{fig:2}). For longer times instead links connect more distant nodes, so that no network property is truly local. On the contrary, as was shown in figure \ref{fig:6}(a), the temporal duration of links depends almost exclusively on the end level, so that near the wall the inner scaling appears.

Instead, at lower Reynolds numbers ($\ret = 180$, 265 and, to a lesser extent, 395) the scaling by outer variables observed at higher Reynolds numbers does not hold. 
While the asymptotic phase of dispersion (approximately $\tau>6$) is the same at all Reynolds numbers (since asymptotic network properties reflect a fully random state), during the transient stage dispersion at low Reynolds numbers appears to be more spatially inhomogeneous and linked to a different underlying mechanism than those bound to the action of outer scales at higher $\ret$. 
While wall-normal velocity fluctuations continue to grow in the Reynolds number interval considered in this work, transport network features, and thus global dispersion properties, achieve Reynolds independence above a certain threshold, which is approximately between $\ret = 395$ and 590. We hypothesize that above this threshold the effects induced by the wall, especially the presence of a region with very low wall-normal velocity variance that traps particles and hence inhibits dispersion, become confined to a portion of the channel half-height $\delta$ that is negligible. Therefore, the contribution of the outer flow in determining the evolution of dispersion becomes dominant, the interaction between cyclic structures and local velocity fluctuations becomes Reynolds-independent, and outer flow scaling is achieved.

\FloatBarrier
\section{Conclusions}
\label{sec:concl}

We analyzed the dispersion of numerically simulated passive tracers in a turbulent channel flow at five different Reynolds numbers, from $\ret = 180$ to $\ret = 950$, by describing the motion of particles with a time-evolving network structure. By doing so, we embedded the high dimensional Lagrangian dynamics of tracers in turbulence into a lower-order representation, which contains and highlights the main features of mixing.
A time-varying complex network was defined to measure the wall-normal transfer of particles between levels of the channel, that is equally spaced partitions of the domain. 
In particular, we focused on the initial transient encountered by dispersing tracers, \textit{i.e.} the period in which their distribution across the height of the channel is still influenced by their release conditions. In this phase, the properties of tracers are highly heterogeneous in the wall-normal direction.

We have demonstrated how the geometrical representation of particle motion can be analyzed from different perspectives, ranging from the quantification of dispersion to the analysis of repeated patterns of motion. These perspectives are an integral part to the complete representation of the features of the flow.
Network-based methods can provide both a reduced-order framework embedding of highly complex fluid dynamics data and the tools to analyze this representation.

We showed the evolution of dispersion for times comparable and longer than the Lagrangian integral timescale and up to reaching asymptotic dispersion, highlighting the wall-normal heterogeneity of the process. We linked the reduced near wall dispersion to two main mechanisms, namely the smaller velocity fluctuations and the presence of cyclic motions. We found that while both mechanisms are present at all the Reynolds numbers considered in this work, the cyclic motions are somewhat more important at the higher Reynolds numbers, while the reduced velocity fluctuations are dominant in inhibiting dispersion at lower $\ret$.
Finally, we showed that, at a sufficiently large Reynolds number, most network quantities are Reynolds independent if scaled with outer flow variables. We linked this behavior to the minor relative importance of near-wall effects and thus to the dominant contribution of the outer flow in driving and influencing dispersion.

Based on the present findings, we showed that the transport network framework, applied to a large set of trajectories integrated for a long time, is able to describe the evolution of mixing at different time scales and to provide new perspectives on the mechanisms involved in the dispersion of particles in an inhomogeneous flow. The transport network is a reduced-order representation of particle motion and its analysis through network-derived tools is able to provide a comprehensive view of dispersion and of its various features. 
It is also worth noting that turbulent flows, despite their very high degree of complexity, can be represented by simpler geometrical objects. In this regard, network-based approaches may provide a foundation upon which reduced-order models of turbulence are built. Finally, we demonstrate the usefulness of this alternative representation by linking the properties of the network to the features of Lagrangian turbulence in wall-bounded flows, showing that the network-based analysis is able to complement classical methods used in the study of inhomogeneous turbulence and extend our knowledge about turbulent mixing.

\FloatBarrier
\begin{acknowledgments}
This work was sponsored by NWO Exacte en Natuurwetenschappen (Physical Sciences) for the use of supercomputer facilities, with financial support from the Netherlands Organization for Scientific Research, NWO. Additional computational resources were provided by HPC@POLITO (\url{https://www.hpc.polito.it})
\end{acknowledgments}

\appendix*
\section{Sensitivity analysis}
\label{sec:sens}

\begin{figure}
\centering
\includegraphics[width = \textwidth]{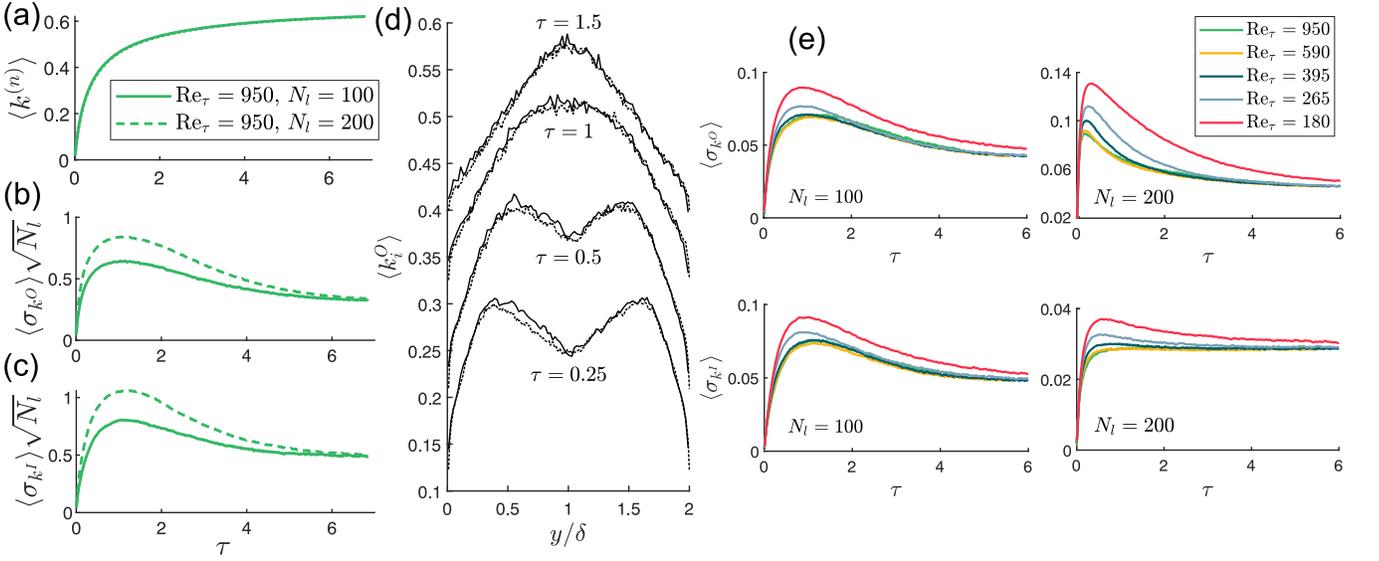}
\caption{(a) Mean degree at $\ret = 950$, simulations with $N_l = 100$, $N_p = \num{10000}$ and $N_l = 200$, $N_p = \num{40000}$. (b) outgoing and (c) ingoing degree standard deviation, normalized by $\sqrt{1/N_l}$, same flow cases as in (a). (d) ensemble averaged outgoing degree versus the $y$ coordinate and at different times (solid lines: $\ret = 950$, $N_l = 100$, $N_p = \num{10000}$; dashed lines:  $\ret = 950$, $N_l = 200$, $N_p = \num{40000}$). (e) outgoing and ingoing degree standard deviation (top and bottom panels, respectively); left panels show all Reynolds numbers with $N_l = 100$ and $N_p = \num{10000}$, while right panels show all Reynolds numbers with $N_l = 200$ and $N_p = \num{10000}$.\label{fig:7}}
\end{figure}
In this appendix the sensitivity of the network representation of dispersion with respect to the number of network levels $N_l$ is analyzed. By doing so, we explore the influence of the spatial scale used to build the network (recall that fluid features smaller than the level size are not explicitly accounted for in the network representation).
To investigate the effects of the number of levels, and thus the spatial size of the discretization employed to build the network, we released another set of $N_b = 60$ batches of particles in the channel flow at $\ret = 950$. Differently from the simulations employed in the main body of the text, we used a release grid composed of $N_l = 200$ levels, each one with 200 particles, for a total of $N_p = \num{40000}$ particles. Keeping the ratio $N_p/N_l^2$ equal to 1 allows us to have the same asymptotic value of the mean degree normalized by $N_l$, \textit{i.e.} $1-1/\mathrm{e}$, found for the $N_l = 100$ case. This property was derived from the assumption that for long times $\tau$ network weights are Poisson distributed \cite{perrone20}; accordingly, the degree is binomially distributed with a success (link existence) probability of $1-1/\mathrm{e}$ and a number of trials equal to $N_l$. 
The mean degree $\langle k^{(n)} \rangle$ is shown in figure \ref{fig:7}(a); the two cases with different $N_l$ exhibit perfect collapse (note that in each case the degree is normalized with the number of levels $N_l$).

Figures \ref{fig:7}(b) and \ref{fig:7}(c) show the values of $\langle \sigma_{k}\rangle$ normalized by $\sqrt{1/N_l}$. Indeed, the asymptotic value of the degree, which is binomially distributed, has a standard deviation proportional to $\sqrt{N_l}$; since we normalize the degree by $N_l$, similarly we divide the degree standard deviations by $\sqrt{1/N_l}$.
Figures 7(b)-(c) show that, before the asymptote, the ingoing and outgoing degree in the case with $N_l = 200$ are more heterogeneous, resulting in larger values of the standard deviations. The increased heterogeneity of the degree is a result of the inclusion of finer scales of motion in the network dynamics. Still, the main features are retained, such as the presence and time of a peak and the imbalance between ingoing and outgoing degree.
Furthermore, the spatial structure of the degree is left mostly unchanged, as can be seen in figure \ref{fig:7}(d), where the ensemble averaged outgoing degree $\langle k_i^{O}\rangle$ is plotted against the channel $y$ coordinate for several times. Differently from figure \ref{fig:3}(a), the mean degree $\langle k^{(n)}\rangle$ was not subtracted to enhance readability.

To analyze the Reynolds number scaling of the network properties and its relationship with the number of levels $N_l$ we used the datasets already employed in the main body of the text, comprising of $N_b$ realizations, each one with $N_p = \num{10000}$ particles. To fill an increased number of levels with particles, we used as the initial time for the network building procedure a time $t_0$ subsequent to the actual release of particles in the channel flow, so that particles have populated all levels. Since in this case (with $N_l>100$) $N_p/N_l^2 \neq 1$, the asymptotic values of the degree statistics are not conserved; furthermore, the degree (especially the outgoing) appears to be more inhomogeneous due to the uneven initial distribution of particles.
However, since in any case the degree of each node is normalized with its maximum attainable value, it is possible to compare the evolution of the degree standard deviation at different Reynolds numbers in the cases with increased $N_l$. Figure \ref{fig:7}(e) shows the ensemble-averaged standard deviations of the outgoing (top panels) and ingoing (bottom panels) degree centralities. In the left panels, the original configuration with $N_l = 100$ computed with the aforementioned procedure is shown. These left panels demonstrate that, similarly to the network generated from the regular grid of particles, curves at $\ret$ greater than 395 collapse exhibiting Reynolds number independence when scaled with outer units. 
Increasing the number of levels to $N_l = 200$ (right panels) leads to the appearance of a slight separation between the degree standard deviation at $\ret = 395$ and those at higher Reynolds numbers, which in turn remain collapsed. We found that increasing the number of levels even further (up to $N_l = 500$, not shown here) a similar structure is retained, with the standard deviations at $\ret = 590$ and 950 collapsed and that at $\ret = 395$ slightly separated. 
The spatial resolution increase of the network introduces the effects of smaller scales of motion, which are particularly important in the near wall region and contribute to the differentiation of the network evolution at lower Reynolds numbers. Still, we found that the scaling at the two higher Reynolds numbers employed in this work is rather insensitive to the increase of the number of spatial levels. Only the point at which the Reynolds independent behavior starts is slightly modified by the increase of resolution.
The sensitivity analysis shows that, even increasing the spatial resolution of the discretization, the main features of dispersion are unchanged. The increase of resolution allows the network representation to account for smaller scales of fluid motion. While this effect is somewhat important at the lower Reynolds numbers, it is negligible at the higher $\ret$ employed here, indicating that outer flow structures are most important in determining the evolution of dispersion.

\bibliography{biblio}

\begin{thebibliography}{54}%
\makeatletter
\providecommand \@ifxundefined [1]{%
 \@ifx{#1\undefined}
}%
\providecommand \@ifnum [1]{%
 \ifnum #1\expandafter \@firstoftwo
 \else \expandafter \@secondoftwo
 \fi
}%
\providecommand \@ifx [1]{%
 \ifx #1\expandafter \@firstoftwo
 \else \expandafter \@secondoftwo
 \fi
}%
\providecommand \natexlab [1]{#1}%
\providecommand \enquote  [1]{``#1''}%
\providecommand \bibnamefont  [1]{#1}%
\providecommand \bibfnamefont [1]{#1}%
\providecommand \citenamefont [1]{#1}%
\providecommand \href@noop [0]{\@secondoftwo}%
\providecommand \href [0]{\begingroup \@sanitize@url \@href}%
\providecommand \@href[1]{\@@startlink{#1}\@@href}%
\providecommand \@@href[1]{\endgroup#1\@@endlink}%
\providecommand \@sanitize@url [0]{\catcode `\\12\catcode `\$12\catcode
  `\&12\catcode `\#12\catcode `\^12\catcode `\_12\catcode `\%12\relax}%
\providecommand \@@startlink[1]{}%
\providecommand \@@endlink[0]{}%
\providecommand \url  [0]{\begingroup\@sanitize@url \@url }%
\providecommand \@url [1]{\endgroup\@href {#1}{\urlprefix }}%
\providecommand \urlprefix  [0]{URL }%
\providecommand \Eprint [0]{\href }%
\providecommand \doibase [0]{https://doi.org/}%
\providecommand \selectlanguage [0]{\@gobble}%
\providecommand \bibinfo  [0]{\@secondoftwo}%
\providecommand \bibfield  [0]{\@secondoftwo}%
\providecommand \translation [1]{[#1]}%
\providecommand \BibitemOpen [0]{}%
\providecommand \bibitemStop [0]{}%
\providecommand \bibitemNoStop [0]{.\EOS\space}%
\providecommand \EOS [0]{\spacefactor3000\relax}%
\providecommand \BibitemShut  [1]{\csname bibitem#1\endcsname}%
\let\auto@bib@innerbib\@empty
\bibitem [{\citenamefont {Dimotakis}(2005)}]{dimotakis2005turbulent}%
  \BibitemOpen
  \bibfield  {author} {\bibinfo {author} {\bibfnamefont {P.~E.}\ \bibnamefont
  {Dimotakis}},\ }\bibfield  {title} {\bibinfo {title} {Turbulent mixing},\
  }\href {https://doi.org/10.1146/annurev.fluid.36.050802.122015} {\bibfield
  {journal} {\bibinfo  {journal} {Annu. Rev. Fluid Mech.}\ }\textbf {\bibinfo
  {volume} {37}},\ \bibinfo {pages} {329} (\bibinfo {year} {2005})}\BibitemShut
  {NoStop}%
\bibitem [{\citenamefont {Toschi}\ and\ \citenamefont
  {Bodenschatz}(2009)}]{toschi2009}%
  \BibitemOpen
  \bibfield  {author} {\bibinfo {author} {\bibfnamefont {F.}~\bibnamefont
  {Toschi}}\ and\ \bibinfo {author} {\bibfnamefont {E.}~\bibnamefont
  {Bodenschatz}},\ }\bibfield  {title} {\bibinfo {title} {Lagrangian
  {{Properties}} of {{Particles}} in {{Turbulence}}},\ }\href
  {https://doi.org/10.1146/annurev.fluid.010908.165210} {\bibfield  {journal}
  {\bibinfo  {journal} {Annual Review of Fluid Mechanics}\ }\textbf {\bibinfo
  {volume} {41}},\ \bibinfo {pages} {375} (\bibinfo {year} {2009})}\BibitemShut
  {NoStop}%
\bibitem [{\citenamefont {Yeung}\ and\ \citenamefont {Pope}(1989)}]{yeung89}%
  \BibitemOpen
  \bibfield  {author} {\bibinfo {author} {\bibfnamefont {P.~K.}\ \bibnamefont
  {Yeung}}\ and\ \bibinfo {author} {\bibfnamefont {S.~B.}\ \bibnamefont
  {Pope}},\ }\bibfield  {title} {\bibinfo {title} {Lagrangian statistics from
  direct numerical simulations of isotropic turbulence},\ }\href
  {https://doi.org/10.1017/S0022112089002697} {\bibfield  {journal} {\bibinfo
  {journal} {Journal of Fluid Mechanics}\ }\textbf {\bibinfo {volume} {207}},\
  \bibinfo {pages} {531} (\bibinfo {year} {1989})}\BibitemShut {NoStop}%
\bibitem [{\citenamefont {Wang}\ \emph {et~al.}(1995)\citenamefont {Wang},
  \citenamefont {Squires},\ and\ \citenamefont {Wu}}]{wang}%
  \BibitemOpen
  \bibfield  {author} {\bibinfo {author} {\bibfnamefont {Q.}~\bibnamefont
  {Wang}}, \bibinfo {author} {\bibfnamefont {K.~D.}\ \bibnamefont {Squires}},\
  and\ \bibinfo {author} {\bibfnamefont {X.}~\bibnamefont {Wu}},\ }\bibfield
  {title} {\bibinfo {title} {Lagrangian statistics in turbulent channel flow},\
  }\href {https://doi.org/https://doi.org/10.1016/1352-2310(95)00190-A}
  {\bibfield  {journal} {\bibinfo  {journal} {Atmospheric Environment}\
  }\textbf {\bibinfo {volume} {29}},\ \bibinfo {pages} {2417} (\bibinfo {year}
  {1995})}\BibitemShut {NoStop}%
\bibitem [{\citenamefont {Toschi}\ \emph {et~al.}(2005)\citenamefont {Toschi},
  \citenamefont {Biferale}, \citenamefont {Boffetta}, \citenamefont {Celani},
  \citenamefont {Devenish},\ and\ \citenamefont {Lanotte}}]{toschi05}%
  \BibitemOpen
  \bibfield  {author} {\bibinfo {author} {\bibfnamefont {F.}~\bibnamefont
  {Toschi}}, \bibinfo {author} {\bibfnamefont {L.}~\bibnamefont {Biferale}},
  \bibinfo {author} {\bibfnamefont {G.}~\bibnamefont {Boffetta}}, \bibinfo
  {author} {\bibfnamefont {A.}~\bibnamefont {Celani}}, \bibinfo {author}
  {\bibfnamefont {B.~J.}\ \bibnamefont {Devenish}},\ and\ \bibinfo {author}
  {\bibfnamefont {A.}~\bibnamefont {Lanotte}},\ }\bibfield  {title} {\bibinfo
  {title} {Acceleration and vortex filaments in turbulence},\ }\href
  {https://doi.org/10.1080/14685240500103150} {\bibfield  {journal} {\bibinfo
  {journal} {Journal of Turbulence}\ }\textbf {\bibinfo {volume} {6}},\
  \bibinfo {pages} {N15} (\bibinfo {year} {2005})}\BibitemShut {NoStop}%
\bibitem [{\citenamefont {Xu}\ \emph {et~al.}(2006)\citenamefont {Xu},
  \citenamefont {Bourgoin}, \citenamefont {Ouellette},\ and\ \citenamefont
  {Bodenschatz}}]{xu06}%
  \BibitemOpen
  \bibfield  {author} {\bibinfo {author} {\bibfnamefont {H.}~\bibnamefont
  {Xu}}, \bibinfo {author} {\bibfnamefont {M.}~\bibnamefont {Bourgoin}},
  \bibinfo {author} {\bibfnamefont {N.~T.}\ \bibnamefont {Ouellette}},\ and\
  \bibinfo {author} {\bibfnamefont {E.}~\bibnamefont {Bodenschatz}},\
  }\bibfield  {title} {\bibinfo {title} {High {{Order Lagrangian Velocity
  Statistics}} in {{Turbulence}}},\ }\href
  {https://doi.org/10.1103/PhysRevLett.96.024503} {\bibfield  {journal}
  {\bibinfo  {journal} {Physical Review Letters}\ }\textbf {\bibinfo {volume}
  {96}},\ \bibinfo {pages} {024503} (\bibinfo {year} {2006})}\BibitemShut
  {NoStop}%
\bibitem [{\citenamefont {Stelzenmuller}\ \emph {et~al.}(2017)\citenamefont
  {Stelzenmuller}, \citenamefont {Polanco}, \citenamefont {Vignal},
  \citenamefont {Vinkovic},\ and\ \citenamefont {Mordant}}]{stelz17}%
  \BibitemOpen
  \bibfield  {author} {\bibinfo {author} {\bibfnamefont {N.}~\bibnamefont
  {Stelzenmuller}}, \bibinfo {author} {\bibfnamefont {J.~I.}\ \bibnamefont
  {Polanco}}, \bibinfo {author} {\bibfnamefont {L.}~\bibnamefont {Vignal}},
  \bibinfo {author} {\bibfnamefont {I.}~\bibnamefont {Vinkovic}},\ and\
  \bibinfo {author} {\bibfnamefont {N.}~\bibnamefont {Mordant}},\ }\bibfield
  {title} {\bibinfo {title} {Lagrangian acceleration statistics in a turbulent
  channel flow},\ }\href {https://doi.org/10.1103/PhysRevFluids.2.054602}
  {\bibfield  {journal} {\bibinfo  {journal} {Physical Review Fluids}\ }\textbf
  {\bibinfo {volume} {2}},\ \bibinfo {pages} {054602} (\bibinfo {year}
  {2017})}\BibitemShut {NoStop}%
\bibitem [{\citenamefont {Salazar}\ and\ \citenamefont
  {Collins}(2009)}]{salazar2009two}%
  \BibitemOpen
  \bibfield  {author} {\bibinfo {author} {\bibfnamefont {J.~P.}\ \bibnamefont
  {Salazar}}\ and\ \bibinfo {author} {\bibfnamefont {L.~R.}\ \bibnamefont
  {Collins}},\ }\bibfield  {title} {\bibinfo {title} {Two-{{Particle
  Dispersion}} in {{Isotropic Turbulent Flows}}},\ }\href
  {https://doi.org/10.1146/annurev.fluid.40.111406.102224} {\bibfield
  {journal} {\bibinfo  {journal} {Annual Review of Fluid Mechanics}\ }\textbf
  {\bibinfo {volume} {41}},\ \bibinfo {pages} {405} (\bibinfo {year}
  {2009})}\BibitemShut {NoStop}%
\bibitem [{\citenamefont {Buaria}\ \emph {et~al.}(2015)\citenamefont {Buaria},
  \citenamefont {Sawford},\ and\ \citenamefont {Yeung}}]{buaria15}%
  \BibitemOpen
  \bibfield  {author} {\bibinfo {author} {\bibfnamefont {D.}~\bibnamefont
  {Buaria}}, \bibinfo {author} {\bibfnamefont {B.~L.}\ \bibnamefont
  {Sawford}},\ and\ \bibinfo {author} {\bibfnamefont {P.~K.}\ \bibnamefont
  {Yeung}},\ }\bibfield  {title} {\bibinfo {title} {Characteristics of backward
  and forward two-particle relative dispersion in turbulence at different
  {{Reynolds}} numbers},\ }\href {https://doi.org/10.1063/1.4931602} {\bibfield
   {journal} {\bibinfo  {journal} {Physics of Fluids}\ }\textbf {\bibinfo
  {volume} {27}},\ \bibinfo {pages} {105101} (\bibinfo {year}
  {2015})}\BibitemShut {NoStop}%
\bibitem [{\citenamefont {Biferale}\ \emph {et~al.}(2005)\citenamefont
  {Biferale}, \citenamefont {Boffetta}, \citenamefont {Celani}, \citenamefont
  {Devenish}, \citenamefont {Lanotte},\ and\ \citenamefont
  {Toschi}}]{biferale2005multiparticle}%
  \BibitemOpen
  \bibfield  {author} {\bibinfo {author} {\bibfnamefont {L.}~\bibnamefont
  {Biferale}}, \bibinfo {author} {\bibfnamefont {G.}~\bibnamefont {Boffetta}},
  \bibinfo {author} {\bibfnamefont {A.}~\bibnamefont {Celani}}, \bibinfo
  {author} {\bibfnamefont {B.~J.}\ \bibnamefont {Devenish}}, \bibinfo {author}
  {\bibfnamefont {A.}~\bibnamefont {Lanotte}},\ and\ \bibinfo {author}
  {\bibfnamefont {F.}~\bibnamefont {Toschi}},\ }\bibfield  {title} {\bibinfo
  {title} {Multiparticle dispersion in fully developed turbulence},\ }\href
  {https://doi.org/10.1063/1.2130751} {\bibfield  {journal} {\bibinfo
  {journal} {Physics of Fluids}\ }\textbf {\bibinfo {volume} {17}},\ \bibinfo
  {pages} {111701} (\bibinfo {year} {2005})}\BibitemShut {NoStop}%
\bibitem [{\citenamefont {Polanco}\ \emph {et~al.}(2018)\citenamefont
  {Polanco}, \citenamefont {Vinkovic}, \citenamefont {Stelzenmuller},
  \citenamefont {Mordant},\ and\ \citenamefont {Bourgoin}}]{polanco18}%
  \BibitemOpen
  \bibfield  {author} {\bibinfo {author} {\bibfnamefont {J.~I.}\ \bibnamefont
  {Polanco}}, \bibinfo {author} {\bibfnamefont {I.}~\bibnamefont {Vinkovic}},
  \bibinfo {author} {\bibfnamefont {N.}~\bibnamefont {Stelzenmuller}}, \bibinfo
  {author} {\bibfnamefont {N.}~\bibnamefont {Mordant}},\ and\ \bibinfo {author}
  {\bibfnamefont {M.}~\bibnamefont {Bourgoin}},\ }\bibfield  {title} {\bibinfo
  {title} {Relative dispersion of particle pairs in turbulent channel flow},\
  }\href {https://doi.org/10.1016/j.ijheatfluidflow.2018.04.007} {\bibfield
  {journal} {\bibinfo  {journal} {International Journal of Heat and Fluid
  Flow}\ }\textbf {\bibinfo {volume} {71}},\ \bibinfo {pages} {231} (\bibinfo
  {year} {2018})}\BibitemShut {NoStop}%
\bibitem [{\citenamefont {Sawford}\ and\ \citenamefont
  {Yeung}(2001)}]{sawford2001stocmodels}%
  \BibitemOpen
  \bibfield  {author} {\bibinfo {author} {\bibfnamefont {B.~L.}\ \bibnamefont
  {Sawford}}\ and\ \bibinfo {author} {\bibfnamefont {P.~K.}\ \bibnamefont
  {Yeung}},\ }\bibfield  {title} {\bibinfo {title} {Lagrangian statistics in
  uniform shear flow: Direct numerical simulation and lagrangian stochastic
  models},\ }\href {https://doi.org/10.1063/1.1388539} {\bibfield  {journal}
  {\bibinfo  {journal} {Physics of Fluids}\ }\textbf {\bibinfo {volume} {13}},\
  \bibinfo {pages} {2627} (\bibinfo {year} {2001})}\BibitemShut {NoStop}%
\bibitem [{\citenamefont {Sawford}(1991)}]{sawford1991pofafd}%
  \BibitemOpen
  \bibfield  {author} {\bibinfo {author} {\bibfnamefont {B.~L.}\ \bibnamefont
  {Sawford}},\ }\bibfield  {title} {\bibinfo {title} {Reynolds number effects
  in {{Lagrangian}} stochastic models of turbulent dispersion},\ }\href
  {https://doi.org/10.1063/1.857937} {\bibfield  {journal} {\bibinfo  {journal}
  {Physics of Fluids A: Fluid Dynamics}\ }\textbf {\bibinfo {volume} {3}},\
  \bibinfo {pages} {1577} (\bibinfo {year} {1991})}\BibitemShut {NoStop}%
\bibitem [{\citenamefont {Smits}\ \emph {et~al.}(2011)\citenamefont {Smits},
  \citenamefont {McKeon},\ and\ \citenamefont {Marusic}}]{smits}%
  \BibitemOpen
  \bibfield  {author} {\bibinfo {author} {\bibfnamefont {A.~J.}\ \bibnamefont
  {Smits}}, \bibinfo {author} {\bibfnamefont {B.~J.}\ \bibnamefont {McKeon}},\
  and\ \bibinfo {author} {\bibfnamefont {I.}~\bibnamefont {Marusic}},\
  }\bibfield  {title} {\bibinfo {title} {High–reynolds number wall
  turbulence},\ }\href {https://doi.org/10.1146/annurev-fluid-122109-160753}
  {\bibfield  {journal} {\bibinfo  {journal} {Annual Review of Fluid
  Mechanics}\ }\textbf {\bibinfo {volume} {43}},\ \bibinfo {pages} {353}
  (\bibinfo {year} {2011})}\BibitemShut {NoStop}%
\bibitem [{\citenamefont {Marusic}\ \emph {et~al.}(2010)\citenamefont
  {Marusic}, \citenamefont {McKeon}, \citenamefont {Monkewitz}, \citenamefont
  {Nagib}, \citenamefont {Smits},\ and\ \citenamefont
  {Sreenivasan}}]{marusic_wb}%
  \BibitemOpen
  \bibfield  {author} {\bibinfo {author} {\bibfnamefont {I.}~\bibnamefont
  {Marusic}}, \bibinfo {author} {\bibfnamefont {B.~J.}\ \bibnamefont {McKeon}},
  \bibinfo {author} {\bibfnamefont {P.~A.}\ \bibnamefont {Monkewitz}}, \bibinfo
  {author} {\bibfnamefont {H.~M.}\ \bibnamefont {Nagib}}, \bibinfo {author}
  {\bibfnamefont {A.~J.}\ \bibnamefont {Smits}},\ and\ \bibinfo {author}
  {\bibfnamefont {K.~R.}\ \bibnamefont {Sreenivasan}},\ }\bibfield  {title}
  {\bibinfo {title} {Wall-bounded turbulent flows at high reynolds numbers:
  Recent advances and key issues},\ }\href {https://doi.org/10.1063/1.3453711}
  {\bibfield  {journal} {\bibinfo  {journal} {Physics of Fluids}\ }\textbf
  {\bibinfo {volume} {22}},\ \bibinfo {pages} {065103} (\bibinfo {year}
  {2010})}\BibitemShut {NoStop}%
\bibitem [{\citenamefont {Iacobello}\ \emph {et~al.}(2021)\citenamefont
  {Iacobello}, \citenamefont {Ridolfi},\ and\ \citenamefont
  {Scarsoglio}}]{iacobelloreview}%
  \BibitemOpen
  \bibfield  {author} {\bibinfo {author} {\bibfnamefont {G.}~\bibnamefont
  {Iacobello}}, \bibinfo {author} {\bibfnamefont {L.}~\bibnamefont {Ridolfi}},\
  and\ \bibinfo {author} {\bibfnamefont {S.}~\bibnamefont {Scarsoglio}},\
  }\bibfield  {title} {\bibinfo {title} {A review on turbulent and vortical
  flow analyses via complex networks},\ }\href
  {https://doi.org/https://doi.org/10.1016/j.physa.2020.125476} {\bibfield
  {journal} {\bibinfo  {journal} {Physica A: Statistical Mechanics and its
  Applications}\ }\textbf {\bibinfo {volume} {563}},\ \bibinfo {pages} {125476}
  (\bibinfo {year} {2021})}\BibitemShut {NoStop}%
\bibitem [{\citenamefont {Donner}\ \emph {et~al.}(2017)\citenamefont {Donner},
  \citenamefont {Hernández-García},\ and\ \citenamefont
  {Ser-Giacomi}}]{donnerreview}%
  \BibitemOpen
  \bibfield  {author} {\bibinfo {author} {\bibfnamefont {R.~V.}\ \bibnamefont
  {Donner}}, \bibinfo {author} {\bibfnamefont {E.}~\bibnamefont
  {Hernández-García}},\ and\ \bibinfo {author} {\bibfnamefont
  {E.}~\bibnamefont {Ser-Giacomi}},\ }\bibfield  {title} {\bibinfo {title}
  {Introduction to focus issue: Complex network perspectives on flow systems},\
  }\href {https://doi.org/10.1063/1.4979129} {\bibfield  {journal} {\bibinfo
  {journal} {Chaos: An Interdisciplinary Journal of Nonlinear Science}\
  }\textbf {\bibinfo {volume} {27}},\ \bibinfo {pages} {035601} (\bibinfo
  {year} {2017})}\BibitemShut {NoStop}%
\bibitem [{\citenamefont {Sujith}\ and\ \citenamefont {Unni}(2020)}]{sujith}%
  \BibitemOpen
  \bibfield  {author} {\bibinfo {author} {\bibfnamefont {R.~I.}\ \bibnamefont
  {Sujith}}\ and\ \bibinfo {author} {\bibfnamefont {V.~R.}\ \bibnamefont
  {Unni}},\ }\bibfield  {title} {\bibinfo {title} {Complex system approach to
  investigate and mitigate thermoacoustic instability in turbulent
  combustors},\ }\href {https://doi.org/10.1063/5.0003702} {\bibfield
  {journal} {\bibinfo  {journal} {Physics of Fluids}\ }\textbf {\bibinfo
  {volume} {32}},\ \bibinfo {pages} {061401} (\bibinfo {year}
  {2020})}\BibitemShut {NoStop}%
\bibitem [{\citenamefont {Yeh}\ \emph {et~al.}(2021)\citenamefont {Yeh},
  \citenamefont {Gopalakrishnan~Meena},\ and\ \citenamefont {Taira}}]{yeh2021}%
  \BibitemOpen
  \bibfield  {author} {\bibinfo {author} {\bibfnamefont {C.-A.}\ \bibnamefont
  {Yeh}}, \bibinfo {author} {\bibfnamefont {M.}~\bibnamefont
  {Gopalakrishnan~Meena}},\ and\ \bibinfo {author} {\bibfnamefont
  {K.}~\bibnamefont {Taira}},\ }\bibfield  {title} {\bibinfo {title} {Network
  broadcast analysis and control of turbulent flows},\ }\href
  {https://doi.org/10.1017/jfm.2020.965} {\bibfield  {journal} {\bibinfo
  {journal} {Journal of Fluid Mechanics}\ }\textbf {\bibinfo {volume} {910}},\
  \bibinfo {pages} {A15} (\bibinfo {year} {2021})}\BibitemShut {NoStop}%
\bibitem [{\citenamefont {Albert}(2005)}]{albert4947}%
  \BibitemOpen
  \bibfield  {author} {\bibinfo {author} {\bibfnamefont {R.}~\bibnamefont
  {Albert}},\ }\bibfield  {title} {\bibinfo {title} {Scale-free networks in
  cell biology},\ }\href {https://doi.org/10.1242/jcs.02714} {\bibfield
  {journal} {\bibinfo  {journal} {Journal of Cell Science}\ }\textbf {\bibinfo
  {volume} {118}},\ \bibinfo {pages} {4947} (\bibinfo {year}
  {2005})}\BibitemShut {NoStop}%
\bibitem [{\citenamefont {Newman}(2018)}]{newman2018networks}%
  \BibitemOpen
  \bibfield  {author} {\bibinfo {author} {\bibfnamefont {M.}~\bibnamefont
  {Newman}},\ }\href@noop {} {\emph {\bibinfo {title} {Networks}}}\ (\bibinfo
  {publisher} {Oxford university press},\ \bibinfo {year} {2018})\BibitemShut
  {NoStop}%
\bibitem [{\citenamefont {Boccaletti}\ \emph {et~al.}(2006)\citenamefont
  {Boccaletti}, \citenamefont {Latora}, \citenamefont {Moreno}, \citenamefont
  {Chavez},\ and\ \citenamefont {Hwang}}]{boccaletti}%
  \BibitemOpen
  \bibfield  {author} {\bibinfo {author} {\bibfnamefont {S.}~\bibnamefont
  {Boccaletti}}, \bibinfo {author} {\bibfnamefont {V.}~\bibnamefont {Latora}},
  \bibinfo {author} {\bibfnamefont {Y.}~\bibnamefont {Moreno}}, \bibinfo
  {author} {\bibfnamefont {M.}~\bibnamefont {Chavez}},\ and\ \bibinfo {author}
  {\bibfnamefont {D.-U.}\ \bibnamefont {Hwang}},\ }\bibfield  {title} {\bibinfo
  {title} {Complex networks: Structure and dynamics},\ }\href
  {https://doi.org/https://doi.org/10.1016/j.physrep.2005.10.009} {\bibfield
  {journal} {\bibinfo  {journal} {Physics Reports}\ }\textbf {\bibinfo {volume}
  {424}},\ \bibinfo {pages} {175 } (\bibinfo {year} {2006})}\BibitemShut
  {NoStop}%
\bibitem [{\citenamefont {Scarsoglio}\ \emph {et~al.}(2016)\citenamefont
  {Scarsoglio}, \citenamefont {Iacobello},\ and\ \citenamefont
  {Ridolfi}}]{scarsoglio2016complex}%
  \BibitemOpen
  \bibfield  {author} {\bibinfo {author} {\bibfnamefont {S.}~\bibnamefont
  {Scarsoglio}}, \bibinfo {author} {\bibfnamefont {G.}~\bibnamefont
  {Iacobello}},\ and\ \bibinfo {author} {\bibfnamefont {L.}~\bibnamefont
  {Ridolfi}},\ }\bibfield  {title} {\bibinfo {title} {Complex networks
  unveiling spatial patterns in turbulence},\ }\href
  {https://doi.org/10.1142/S0218127416502230} {\bibfield  {journal} {\bibinfo
  {journal} {International Journal of Bifurcation and Chaos}\ }\textbf
  {\bibinfo {volume} {26}},\ \bibinfo {pages} {1650223} (\bibinfo {year}
  {2016})}\BibitemShut {NoStop}%
\bibitem [{\citenamefont {Schlueter-Kuck}\ and\ \citenamefont
  {Dabiri}(2017)}]{schlueter2017coherent}%
  \BibitemOpen
  \bibfield  {author} {\bibinfo {author} {\bibfnamefont {K.~L.}\ \bibnamefont
  {Schlueter-Kuck}}\ and\ \bibinfo {author} {\bibfnamefont {J.~O.}\
  \bibnamefont {Dabiri}},\ }\bibfield  {title} {\bibinfo {title} {Coherent
  structure colouring: identification of coherent structures from sparse data
  using graph theory},\ }\href {https://doi.org/10.1017/jfm.2016.755}
  {\bibfield  {journal} {\bibinfo  {journal} {Journal of Fluid Mechanics}\
  }\textbf {\bibinfo {volume} {811}},\ \bibinfo {pages} {468} (\bibinfo {year}
  {2017})}\BibitemShut {NoStop}%
\bibitem [{\citenamefont {Vieweg}\ \emph {et~al.}(2021)\citenamefont {Vieweg},
  \citenamefont {Schneide}, \citenamefont {Padberg-Gehle},\ and\ \citenamefont
  {Schumacher}}]{vieweg}%
  \BibitemOpen
  \bibfield  {author} {\bibinfo {author} {\bibfnamefont {P.~P.}\ \bibnamefont
  {Vieweg}}, \bibinfo {author} {\bibfnamefont {C.}~\bibnamefont {Schneide}},
  \bibinfo {author} {\bibfnamefont {K.}~\bibnamefont {Padberg-Gehle}},\ and\
  \bibinfo {author} {\bibfnamefont {J.}~\bibnamefont {Schumacher}},\ }\bibfield
   {title} {\bibinfo {title} {Lagrangian heat transport in turbulent
  three-dimensional convection},\ }\href
  {https://doi.org/10.1103/PhysRevFluids.6.L041501} {\bibfield  {journal}
  {\bibinfo  {journal} {Phys. Rev. Fluids}\ }\textbf {\bibinfo {volume} {6}},\
  \bibinfo {pages} {L041501} (\bibinfo {year} {2021})}\BibitemShut {NoStop}%
\bibitem [{\citenamefont {Charakopoulos}\ \emph {et~al.}(2014)\citenamefont
  {Charakopoulos}, \citenamefont {Karakasidis}, \citenamefont {Papanicolaou},\
  and\ \citenamefont {Liakopoulos}}]{charakopoulos}%
  \BibitemOpen
  \bibfield  {author} {\bibinfo {author} {\bibfnamefont {A.~K.}\ \bibnamefont
  {Charakopoulos}}, \bibinfo {author} {\bibfnamefont {T.~E.}\ \bibnamefont
  {Karakasidis}}, \bibinfo {author} {\bibfnamefont {P.~N.}\ \bibnamefont
  {Papanicolaou}},\ and\ \bibinfo {author} {\bibfnamefont {A.}~\bibnamefont
  {Liakopoulos}},\ }\bibfield  {title} {\bibinfo {title} {The application of
  complex network time series analysis in turbulent heated jets},\ }\href
  {https://doi.org/10.1063/1.4875040} {\bibfield  {journal} {\bibinfo
  {journal} {Chaos: An Interdisciplinary Journal of Nonlinear Science}\
  }\textbf {\bibinfo {volume} {24}},\ \bibinfo {pages} {024408} (\bibinfo
  {year} {2014})}\BibitemShut {NoStop}%
\bibitem [{\citenamefont {Charakopoulos}\ \emph {et~al.}(2021)\citenamefont
  {Charakopoulos}, \citenamefont {Karakasidis},\ and\ \citenamefont
  {Sarris}}]{chara21}%
  \BibitemOpen
  \bibfield  {author} {\bibinfo {author} {\bibfnamefont {A.}~\bibnamefont
  {Charakopoulos}}, \bibinfo {author} {\bibfnamefont {T.}~\bibnamefont
  {Karakasidis}},\ and\ \bibinfo {author} {\bibfnamefont {I.}~\bibnamefont
  {Sarris}},\ }\bibfield  {title} {\bibinfo {title} {Analysis of
  magnetohydrodynamic channel flow through complex network analysis},\ }\href
  {https://doi.org/10.1063/5.0043817} {\bibfield  {journal} {\bibinfo
  {journal} {Chaos: An Interdisciplinary Journal of Nonlinear Science}\
  }\textbf {\bibinfo {volume} {31}},\ \bibinfo {pages} {043123} (\bibinfo
  {year} {2021})}\BibitemShut {NoStop}%
\bibitem [{\citenamefont {Rypina}\ and\ \citenamefont
  {Pratt}(2017)}]{rypina2017npg}%
  \BibitemOpen
  \bibfield  {author} {\bibinfo {author} {\bibfnamefont {I.~I.}\ \bibnamefont
  {Rypina}}\ and\ \bibinfo {author} {\bibfnamefont {L.~J.}\ \bibnamefont
  {Pratt}},\ }\bibfield  {title} {\bibinfo {title} {Trajectory encounter volume
  as a diagnostic of mixing potential in fluid flows},\ }\href
  {https://doi.org/10.5194/npg-24-189-2017} {\bibfield  {journal} {\bibinfo
  {journal} {Nonlinear Processes in Geophysics}\ }\textbf {\bibinfo {volume}
  {24}},\ \bibinfo {pages} {189} (\bibinfo {year} {2017})}\BibitemShut
  {NoStop}%
\bibitem [{\citenamefont {Banisch}\ \emph {et~al.}(2019)\citenamefont
  {Banisch}, \citenamefont {Koltai},\ and\ \citenamefont
  {Padberg-Gehle}}]{banisch}%
  \BibitemOpen
  \bibfield  {author} {\bibinfo {author} {\bibfnamefont {R.}~\bibnamefont
  {Banisch}}, \bibinfo {author} {\bibfnamefont {P.}~\bibnamefont {Koltai}},\
  and\ \bibinfo {author} {\bibfnamefont {K.}~\bibnamefont {Padberg-Gehle}},\
  }\bibfield  {title} {\bibinfo {title} {Network measures of mixing},\ }\href
  {https://doi.org/10.1063/1.5087632} {\bibfield  {journal} {\bibinfo
  {journal} {Chaos: An Interdisciplinary Journal of Nonlinear Science}\
  }\textbf {\bibinfo {volume} {29}},\ \bibinfo {pages} {063125} (\bibinfo
  {year} {2019})}\BibitemShut {NoStop}%
\bibitem [{\citenamefont {Taira}\ \emph {et~al.}(2016)\citenamefont {Taira},
  \citenamefont {Nair},\ and\ \citenamefont {Brunton}}]{taira2016jfm}%
  \BibitemOpen
  \bibfield  {author} {\bibinfo {author} {\bibfnamefont {K.}~\bibnamefont
  {Taira}}, \bibinfo {author} {\bibfnamefont {A.~G.}\ \bibnamefont {Nair}},\
  and\ \bibinfo {author} {\bibfnamefont {S.~L.}\ \bibnamefont {Brunton}},\
  }\bibfield  {title} {\bibinfo {title} {Network structure of two-dimensional
  decaying isotropic turbulence},\ }\bibfield  {journal} {\bibinfo  {journal}
  {Journal of Fluid Mechanics}\ }\textbf {\bibinfo {volume} {795}},\ \href
  {https://doi.org/10.1017/jfm.2016.235} {10.1017/jfm.2016.235} (\bibinfo
  {year} {2016})\BibitemShut {NoStop}%
\bibitem [{\citenamefont {Froyland}\ \emph {et~al.}(2007)\citenamefont
  {Froyland}, \citenamefont {Padberg}, \citenamefont {England},\ and\
  \citenamefont {Treguier}}]{froyland2007prl}%
  \BibitemOpen
  \bibfield  {author} {\bibinfo {author} {\bibfnamefont {G.}~\bibnamefont
  {Froyland}}, \bibinfo {author} {\bibfnamefont {K.}~\bibnamefont {Padberg}},
  \bibinfo {author} {\bibfnamefont {M.~H.}\ \bibnamefont {England}},\ and\
  \bibinfo {author} {\bibfnamefont {A.~M.}\ \bibnamefont {Treguier}},\
  }\bibfield  {title} {\bibinfo {title} {Detection of coherent oceanic
  structures via transfer operators},\ }\href
  {https://doi.org/10.1103/PhysRevLett.98.224503} {\bibfield  {journal}
  {\bibinfo  {journal} {Phys. Rev. Lett.}\ }\textbf {\bibinfo {volume} {98}},\
  \bibinfo {pages} {224503} (\bibinfo {year} {2007})}\BibitemShut {NoStop}%
\bibitem [{\citenamefont {Ser-Giacomi}\ \emph {et~al.}(2015)\citenamefont
  {Ser-Giacomi}, \citenamefont {Rossi}, \citenamefont {L{\'o}pez},\ and\
  \citenamefont {Hernandez-Garcia}}]{sergiac}%
  \BibitemOpen
  \bibfield  {author} {\bibinfo {author} {\bibfnamefont {E.}~\bibnamefont
  {Ser-Giacomi}}, \bibinfo {author} {\bibfnamefont {V.}~\bibnamefont {Rossi}},
  \bibinfo {author} {\bibfnamefont {C.}~\bibnamefont {L{\'o}pez}},\ and\
  \bibinfo {author} {\bibfnamefont {E.}~\bibnamefont {Hernandez-Garcia}},\
  }\bibfield  {title} {\bibinfo {title} {Flow networks: A characterization of
  geophysical fluid transport},\ }\href
  {https://doi.org/https://doi.org/10.1063/1.4908231} {\bibfield  {journal}
  {\bibinfo  {journal} {Chaos: An Interdisciplinary Journal of Nonlinear
  Science}\ }\textbf {\bibinfo {volume} {25}},\ \bibinfo {pages} {036404}
  (\bibinfo {year} {2015})}\BibitemShut {NoStop}%
\bibitem [{\citenamefont {Iacobello}\ \emph
  {et~al.}(2019{\natexlab{a}})\citenamefont {Iacobello}, \citenamefont {Marro},
  \citenamefont {Ridolfi}, \citenamefont {Salizzoni},\ and\ \citenamefont
  {Scarsoglio}}]{iacobello_plume}%
  \BibitemOpen
  \bibfield  {author} {\bibinfo {author} {\bibfnamefont {G.}~\bibnamefont
  {Iacobello}}, \bibinfo {author} {\bibfnamefont {M.}~\bibnamefont {Marro}},
  \bibinfo {author} {\bibfnamefont {L.}~\bibnamefont {Ridolfi}}, \bibinfo
  {author} {\bibfnamefont {P.}~\bibnamefont {Salizzoni}},\ and\ \bibinfo
  {author} {\bibfnamefont {S.}~\bibnamefont {Scarsoglio}},\ }\bibfield  {title}
  {\bibinfo {title} {Experimental investigation of vertical turbulent transport
  of a passive scalar in a boundary layer: Statistics and visibility graph
  analysis},\ }\href {https://doi.org/10.1103/PhysRevFluids.4.104501}
  {\bibfield  {journal} {\bibinfo  {journal} {Phys. Rev. Fluids}\ }\textbf
  {\bibinfo {volume} {4}},\ \bibinfo {pages} {104501} (\bibinfo {year}
  {2019}{\natexlab{a}})}\BibitemShut {NoStop}%
\bibitem [{\citenamefont {Iacobello}\ \emph
  {et~al.}(2019{\natexlab{b}})\citenamefont {Iacobello}, \citenamefont
  {Scarsoglio}, \citenamefont {Kuerten},\ and\ \citenamefont
  {Ridolfi}}]{iacobello2019lagrangian}%
  \BibitemOpen
  \bibfield  {author} {\bibinfo {author} {\bibfnamefont {G.}~\bibnamefont
  {Iacobello}}, \bibinfo {author} {\bibfnamefont {S.}~\bibnamefont
  {Scarsoglio}}, \bibinfo {author} {\bibfnamefont {J.}~\bibnamefont
  {Kuerten}},\ and\ \bibinfo {author} {\bibfnamefont {L.}~\bibnamefont
  {Ridolfi}},\ }\bibfield  {title} {\bibinfo {title} {Lagrangian network
  analysis of turbulent mixing},\ }\href {https://doi.org/10.1017/jfm.2019.79}
  {\bibfield  {journal} {\bibinfo  {journal} {Journal of Fluid Mechanics}\
  }\textbf {\bibinfo {volume} {865}},\ \bibinfo {pages} {546} (\bibinfo {year}
  {2019}{\natexlab{b}})}\BibitemShut {NoStop}%
\bibitem [{\citenamefont {Perrone}\ \emph {et~al.}(2020)\citenamefont
  {Perrone}, \citenamefont {Kuerten}, \citenamefont {Ridolfi},\ and\
  \citenamefont {Scarsoglio}}]{perrone20}%
  \BibitemOpen
  \bibfield  {author} {\bibinfo {author} {\bibfnamefont {D.}~\bibnamefont
  {Perrone}}, \bibinfo {author} {\bibfnamefont {J.~G.~M.}\ \bibnamefont
  {Kuerten}}, \bibinfo {author} {\bibfnamefont {L.}~\bibnamefont {Ridolfi}},\
  and\ \bibinfo {author} {\bibfnamefont {S.}~\bibnamefont {Scarsoglio}},\
  }\bibfield  {title} {\bibinfo {title} {Wall-induced anisotropy effects on
  turbulent mixing in channel flow: A network-based analysis},\ }\href
  {https://doi.org/10.1103/PhysRevE.102.043109} {\bibfield  {journal} {\bibinfo
   {journal} {Phys. Rev. E}\ }\textbf {\bibinfo {volume} {102}},\ \bibinfo
  {pages} {043109} (\bibinfo {year} {2020})}\BibitemShut {NoStop}%
\bibitem [{\citenamefont {Kim}\ \emph {et~al.}(1987)\citenamefont {Kim},
  \citenamefont {Moin},\ and\ \citenamefont {Moser}}]{kim_moin_moser_1987}%
  \BibitemOpen
  \bibfield  {author} {\bibinfo {author} {\bibfnamefont {J.}~\bibnamefont
  {Kim}}, \bibinfo {author} {\bibfnamefont {P.}~\bibnamefont {Moin}},\ and\
  \bibinfo {author} {\bibfnamefont {R.}~\bibnamefont {Moser}},\ }\bibfield
  {title} {\bibinfo {title} {Turbulence statistics in fully developed channel
  flow at low reynolds number},\ }\href
  {https://doi.org/10.1017/S0022112087000892} {\bibfield  {journal} {\bibinfo
  {journal} {Journal of Fluid Mechanics}\ }\textbf {\bibinfo {volume} {177}},\
  \bibinfo {pages} {133–166} (\bibinfo {year} {1987})}\BibitemShut {NoStop}%
\bibitem [{\citenamefont {Canuto}\ \emph {et~al.}(2012)\citenamefont {Canuto},
  \citenamefont {Hussaini}, \citenamefont {Quarteroni}, \citenamefont
  {Thomas~Jr} \emph {et~al.}}]{canuto2012spectral}%
  \BibitemOpen
  \bibfield  {author} {\bibinfo {author} {\bibfnamefont {C.}~\bibnamefont
  {Canuto}}, \bibinfo {author} {\bibfnamefont {M.~Y.}\ \bibnamefont
  {Hussaini}}, \bibinfo {author} {\bibfnamefont {A.}~\bibnamefont
  {Quarteroni}}, \bibinfo {author} {\bibfnamefont {A.}~\bibnamefont
  {Thomas~Jr}}, \emph {et~al.},\ }\href@noop {} {\emph {\bibinfo {title}
  {Spectral methods in fluid dynamics}}}\ (\bibinfo  {publisher} {Springer
  Science \& Business Media},\ \bibinfo {year} {2012})\BibitemShut {NoStop}%
\bibitem [{\citenamefont {Kuerten}\ and\ \citenamefont
  {Brouwers}(2013)}]{kuerten13}%
  \BibitemOpen
  \bibfield  {author} {\bibinfo {author} {\bibfnamefont {J.~G.}\ \bibnamefont
  {Kuerten}}\ and\ \bibinfo {author} {\bibfnamefont {J.}~\bibnamefont
  {Brouwers}},\ }\bibfield  {title} {\bibinfo {title} {Lagrangian statistics of
  turbulent channel flow at $\mathrm{Re}_{\tau}$= 950 calculated with direct
  numerical simulation and langevin models},\ }\href
  {https://doi.org/https://doi.org/10.1063/1.4824795} {\bibfield  {journal}
  {\bibinfo  {journal} {Physics of fluids}\ }\textbf {\bibinfo {volume} {25}},\
  \bibinfo {pages} {105108} (\bibinfo {year} {2013})}\BibitemShut {NoStop}%
\bibitem [{\citenamefont {Perrone}\ \emph {et~al.}(2021)\citenamefont
  {Perrone}, \citenamefont {Kuerten}, \citenamefont {Ridolfi},\ and\
  \citenamefont {Scarsoglio}}]{lagstat}%
  \BibitemOpen
  \bibfield  {author} {\bibinfo {author} {\bibfnamefont {D.}~\bibnamefont
  {Perrone}}, \bibinfo {author} {\bibfnamefont {J.~G.~M.}\ \bibnamefont
  {Kuerten}}, \bibinfo {author} {\bibfnamefont {L.}~\bibnamefont {Ridolfi}},\
  and\ \bibinfo {author} {\bibfnamefont {S.}~\bibnamefont {Scarsoglio}},\
  }\bibfield  {title} {\bibinfo {title} {Lagrangian statistics in turbulent
  channel flow},\ }\href {https://doi.org/10.5281/zenodo.4916024}
  {10.5281/zenodo.4916024} (\bibinfo {year} {2021})\BibitemShut {NoStop}%
\bibitem [{\citenamefont {Taylor}(1922)}]{taylor_diff}%
  \BibitemOpen
  \bibfield  {author} {\bibinfo {author} {\bibfnamefont {G.~I.}\ \bibnamefont
  {Taylor}},\ }\bibfield  {title} {\bibinfo {title} {Diffusion by continuous
  movements},\ }\href@noop {} {\bibfield  {journal} {\bibinfo  {journal}
  {Proceedings of the London mathematical society}\ }\textbf {\bibinfo {volume}
  {2}},\ \bibinfo {pages} {196} (\bibinfo {year} {1922})}\BibitemShut {NoStop}%
\bibitem [{\citenamefont {Stelzenmuller}(2017)}]{stelzenmuller2017lagrangian}%
  \BibitemOpen
  \bibfield  {author} {\bibinfo {author} {\bibfnamefont {N.}~\bibnamefont
  {Stelzenmuller}},\ }\emph {\bibinfo {title} {A Lagrangian study of
  inhomogeneous turbulence}},\ \href
  {https://tel.archives-ouvertes.fr/tel-01739689/document} {Ph.D. thesis}
  (\bibinfo {year} {2017})\BibitemShut {NoStop}%
\bibitem [{\citenamefont {Kruskal}\ and\ \citenamefont
  {Wallis}(1952)}]{kruskalwallis}%
  \BibitemOpen
  \bibfield  {author} {\bibinfo {author} {\bibfnamefont {W.~H.}\ \bibnamefont
  {Kruskal}}\ and\ \bibinfo {author} {\bibfnamefont {W.~A.}\ \bibnamefont
  {Wallis}},\ }\bibfield  {title} {\bibinfo {title} {Use of ranks in
  one-criterion variance analysis},\ }\href
  {http://www.jstor.org/stable/2280779} {\bibfield  {journal} {\bibinfo
  {journal} {Journal of the American Statistical Association}\ }\textbf
  {\bibinfo {volume} {47}},\ \bibinfo {pages} {583} (\bibinfo {year}
  {1952})}\BibitemShut {NoStop}%
\bibitem [{\citenamefont {Jucha}\ \emph {et~al.}(2014)\citenamefont {Jucha},
  \citenamefont {Xu}, \citenamefont {Pumir},\ and\ \citenamefont
  {Bodenschatz}}]{jucha14}%
  \BibitemOpen
  \bibfield  {author} {\bibinfo {author} {\bibfnamefont {J.}~\bibnamefont
  {Jucha}}, \bibinfo {author} {\bibfnamefont {H.}~\bibnamefont {Xu}}, \bibinfo
  {author} {\bibfnamefont {A.}~\bibnamefont {Pumir}},\ and\ \bibinfo {author}
  {\bibfnamefont {E.}~\bibnamefont {Bodenschatz}},\ }\bibfield  {title}
  {\bibinfo {title} {Time-reversal-symmetry breaking in turbulence},\ }\href
  {https://doi.org/10.1103/PhysRevLett.113.054501} {\bibfield  {journal}
  {\bibinfo  {journal} {Phys. Rev. Lett.}\ }\textbf {\bibinfo {volume} {113}},\
  \bibinfo {pages} {054501} (\bibinfo {year} {2014})}\BibitemShut {NoStop}%
\bibitem [{\citenamefont {Bragg}\ \emph {et~al.}(2016)\citenamefont {Bragg},
  \citenamefont {Ireland},\ and\ \citenamefont {Collins}}]{bragg}%
  \BibitemOpen
  \bibfield  {author} {\bibinfo {author} {\bibfnamefont {A.~D.}\ \bibnamefont
  {Bragg}}, \bibinfo {author} {\bibfnamefont {P.~J.}\ \bibnamefont {Ireland}},\
  and\ \bibinfo {author} {\bibfnamefont {L.~R.}\ \bibnamefont {Collins}},\
  }\bibfield  {title} {\bibinfo {title} {Forward and backward in time
  dispersion of fluid and inertial particles in isotropic turbulence},\ }\href
  {https://doi.org/10.1063/1.4939694} {\bibfield  {journal} {\bibinfo
  {journal} {Physics of Fluids}\ }\textbf {\bibinfo {volume} {28}},\ \bibinfo
  {pages} {013305} (\bibinfo {year} {2016})}\BibitemShut {NoStop}%
\bibitem [{\citenamefont {Polanco}(2019)}]{polanco_thesis}%
  \BibitemOpen
  \bibfield  {author} {\bibinfo {author} {\bibfnamefont {J.~I.}\ \bibnamefont
  {Polanco}},\ }\emph {\bibinfo {title} {{Lagrangian properties of turbulent
  channel flow : a numerical study}}},\ \href
  {https://hal.archives-ouvertes.fr/tel-02084215} {\bibinfo {type} {Theses}},\
  \bibinfo  {school} {{Universit{\'e} de Lyon}} (\bibinfo {year}
  {2019})\BibitemShut {NoStop}%
\bibitem [{\citenamefont {Rodríguez-Méndez}\ \emph
  {et~al.}(2017)\citenamefont {Rodríguez-Méndez}, \citenamefont
  {Ser-Giacomi},\ and\ \citenamefont {Hernández-García}}]{rodriguez}%
  \BibitemOpen
  \bibfield  {author} {\bibinfo {author} {\bibfnamefont {V.}~\bibnamefont
  {Rodríguez-Méndez}}, \bibinfo {author} {\bibfnamefont {E.}~\bibnamefont
  {Ser-Giacomi}},\ and\ \bibinfo {author} {\bibfnamefont {E.}~\bibnamefont
  {Hernández-García}},\ }\bibfield  {title} {\bibinfo {title} {Clustering
  coefficient and periodic orbits in flow networks},\ }\href
  {https://doi.org/10.1063/1.4971787} {\bibfield  {journal} {\bibinfo
  {journal} {Chaos: An Interdisciplinary Journal of Nonlinear Science}\
  }\textbf {\bibinfo {volume} {27}},\ \bibinfo {pages} {035803} (\bibinfo
  {year} {2017})}\BibitemShut {NoStop}%
\bibitem [{\citenamefont {Schwarze}\ and\ \citenamefont
  {Porter}(2020)}]{schwarze2020motifs}%
  \BibitemOpen
  \bibfield  {author} {\bibinfo {author} {\bibfnamefont {A.~C.}\ \bibnamefont
  {Schwarze}}\ and\ \bibinfo {author} {\bibfnamefont {M.~A.}\ \bibnamefont
  {Porter}},\ }\href@noop {} {\bibinfo {title} {Motifs for processes on
  networks}} (\bibinfo {year} {2020}),\ \Eprint
  {https://arxiv.org/abs/2007.07447} {arXiv:2007.07447 [physics.soc-ph]}
  \BibitemShut {NoStop}%
\bibitem [{\citenamefont {Alon}(2007)}]{alon2007network}%
  \BibitemOpen
  \bibfield  {author} {\bibinfo {author} {\bibfnamefont {U.}~\bibnamefont
  {Alon}},\ }\bibfield  {title} {\bibinfo {title} {Network motifs: theory and
  experimental approaches},\ }\href {https://doi.org/10.1038/nrg2102}
  {\bibfield  {journal} {\bibinfo  {journal} {Nature Reviews Genetics}\
  }\textbf {\bibinfo {volume} {8}},\ \bibinfo {pages} {450} (\bibinfo {year}
  {2007})}\BibitemShut {NoStop}%
\bibitem [{\citenamefont {{Wernicke}}(2006)}]{wernicke}%
  \BibitemOpen
  \bibfield  {author} {\bibinfo {author} {\bibfnamefont {S.}~\bibnamefont
  {{Wernicke}}},\ }\bibfield  {title} {\bibinfo {title} {Efficient detection of
  network motifs},\ }\href {https://doi.org/10.1109/TCBB.2006.51} {\bibfield
  {journal} {\bibinfo  {journal} {IEEE/ACM Transactions on Computational
  Biology and Bioinformatics}\ }\textbf {\bibinfo {volume} {3}},\ \bibinfo
  {pages} {347} (\bibinfo {year} {2006})}\BibitemShut {NoStop}%
\bibitem [{\citenamefont {Fagiolo}(2007)}]{fagiolo}%
  \BibitemOpen
  \bibfield  {author} {\bibinfo {author} {\bibfnamefont {G.}~\bibnamefont
  {Fagiolo}},\ }\bibfield  {title} {\bibinfo {title} {Clustering in complex
  directed networks},\ }\href {https://doi.org/10.1103/PhysRevE.76.026107}
  {\bibfield  {journal} {\bibinfo  {journal} {Phys. Rev. E}\ }\textbf {\bibinfo
  {volume} {76}},\ \bibinfo {pages} {026107} (\bibinfo {year}
  {2007})}\BibitemShut {NoStop}%
\bibitem [{\citenamefont {Cencini}\ \emph {et~al.}(2009)\citenamefont
  {Cencini}, \citenamefont {Cecconi},\ and\ \citenamefont {Vulpiani}}]{chaos}%
  \BibitemOpen
  \bibfield  {author} {\bibinfo {author} {\bibfnamefont {M.}~\bibnamefont
  {Cencini}}, \bibinfo {author} {\bibfnamefont {F.}~\bibnamefont {Cecconi}},\
  and\ \bibinfo {author} {\bibfnamefont {A.}~\bibnamefont {Vulpiani}},\ }\href
  {https://doi.org/10.1142/7351} {\emph {\bibinfo {title} {Chaos}}}\ (\bibinfo
  {publisher} {WORLD SCIENTIFIC},\ \bibinfo {year} {2009})\BibitemShut
  {NoStop}%
\bibitem [{\citenamefont {Bonacich}(1972)}]{bonacich}%
  \BibitemOpen
  \bibfield  {author} {\bibinfo {author} {\bibfnamefont {P.}~\bibnamefont
  {Bonacich}},\ }\bibfield  {title} {\bibinfo {title} {Factoring and weighting
  approaches to status scores and clique identification},\ }\href
  {https://doi.org/10.1080/0022250X.1972.9989806} {\bibfield  {journal}
  {\bibinfo  {journal} {The Journal of Mathematical Sociology}\ }\textbf
  {\bibinfo {volume} {2}},\ \bibinfo {pages} {113} (\bibinfo {year}
  {1972})}\BibitemShut {NoStop}%
\bibitem [{\citenamefont {Froyland}\ and\ \citenamefont
  {Padberg}(2009)}]{froyland09}%
  \BibitemOpen
  \bibfield  {author} {\bibinfo {author} {\bibfnamefont {G.}~\bibnamefont
  {Froyland}}\ and\ \bibinfo {author} {\bibfnamefont {K.}~\bibnamefont
  {Padberg}},\ }\bibfield  {title} {\bibinfo {title} {Almost-invariant sets and
  invariant manifolds — connecting probabilistic and geometric descriptions
  of coherent structures in flows},\ }\href
  {https://doi.org/https://doi.org/10.1016/j.physd.2009.03.002} {\bibfield
  {journal} {\bibinfo  {journal} {Physica D: Nonlinear Phenomena}\ }\textbf
  {\bibinfo {volume} {238}},\ \bibinfo {pages} {1507} (\bibinfo {year}
  {2009})}\BibitemShut {NoStop}%
\bibitem [{\citenamefont {Sawford}(1986)}]{sawford_wm}%
  \BibitemOpen
  \bibfield  {author} {\bibinfo {author} {\bibfnamefont {B.~L.}\ \bibnamefont
  {Sawford}},\ }\bibfield  {title} {\bibinfo {title} {Generalized random
  forcing in random‐walk turbulent dispersion models},\ }\href
  {https://doi.org/10.1063/1.865784} {\bibfield  {journal} {\bibinfo  {journal}
  {The Physics of Fluids}\ }\textbf {\bibinfo {volume} {29}},\ \bibinfo {pages}
  {3582} (\bibinfo {year} {1986})}\BibitemShut {NoStop}%
\end{thebibliography}%

\end{document}